\documentclass[11pt]{article}
\usepackage{amsmath,amssymb,bm,epsf,epsfig,graphicx}
\input epsf.sty
\usepackage[utf8]{inputenc}
\usepackage[english]{babel}
\usepackage{amsmath}
\usepackage{latexsym}
\usepackage{amsfonts}
\usepackage{mathtools}
\usepackage{amssymb}
\usepackage{scalerel}
\usepackage{extpfeil}
\usepackage[left=3cm,right=3cm,top=3.1cm,bottom=3.1cm]{geometry}
\usepackage{cite}
\usepackage{tensor}
\usepackage{tikz}
\usepackage{tikz-cd}
\usepackage[mathscr]{euscript}
\usetikzlibrary{arrows.meta, bending}
\tikzset{C/.style={circle, minimum size=8mm,
		node contents={},
		append after command={\pgfextra{%
				\draw[-{Straight Barb[flex']}](\tikzlastnode.150) arc (450:110:2.8mm);}
	}}
}

\newcommand{\bs}[1]{\boldsymbol{#1}}

\usepackage{hyperref}
\numberwithin{equation}{section}

\newcommand{\ad}{\mathop{\rm ad}\nolimits}

\def\ket#1{|#1 \rangle}
\def\aver#1{\left\langle\, #1 \,\right\rangle}

\def\p{\partial}

\def \be {\begin{eqnarray}}
\def \ee {\end{eqnarray}}
\def \bal {\begin{align}}
\def \eal {\end{align}}
\def \bdm {\begin{displaymath}}
\def \edm {\end{displaymath}}

\def \tr{{\rm tr}}

\def\del {\partial}
\def\0{\nonumber}

\begin{document}
	\begingroup\allowdisplaybreaks

\vspace*{1.1cm}

\centerline{\Large \bf Closed string deformations  in open string field theory II: }\vspace{.3cm}
 \centerline{\Large \bf Superstring}
\vspace*{.1cm}

\begin{center}

{\large Carlo Maccaferri$^{(a)}$\footnote{Email: maccafer at gmail.com} and Jakub Vo\v{s}mera$^{(b,c)}$\footnote{Email: jvosmera at phys.ethz.ch} }
\vskip 1 cm
$^{(a)}${\it Dipartimento di Fisica, Universit\`a di Torino, \\INFN  Sezione di Torino and Arnold-Regge Center\\
Via Pietro Giuria 1, I-10125 Torino, Italy}
\vskip .5 cm
$^{(b)}${\it Institut f\"{u}r Theoretische Physik, ETH Z\"{u}rich\\
	Wolfgang-Pauli-Straße 27, 8093 Z\"{u}rich, Switzerland}

\end{center}

\vspace*{6.0ex}

\centerline{\bf Abstract}
\bigskip
This is the second paper of a series of three. We construct effective open-closed superstring couplings by classically integrating out massive fields from open superstring field theories coupled to an elementary gauge invariant tadpole proportional to an on-shell closed string state in both large and small Hilbert spaces, in the NS sector. This source term is well known in the WZW formulation  and by explicitly performing a novel large Hilbert space perturbation theory  we are able to characterize the first orders of the vacuum shift solution, its obstructions and the non-trivial open-closed effective couplings in closed form. With the aim of getting all order results, we also construct a new observable in the $A_\infty$ theory in the small Hilbert space which correctly provides a gauge invariant coupling to physical closed strings and which descends from the WZW open-closed coupling upon partial gauge fixing and field redefinition.  Armed with this new $A_\infty$ observable  we use tensor co-algebra techniques to efficiently package the whole perturbation theory necessary for computing the effective action  and we give all order results for the open-closed effective couplings in the small Hilbert space.

\baselineskip=16pt
\newpage
\tableofcontents

\section{Introduction and summary}
This second paper of the series including \cite{OC-I,OC-III} is devoted to the coupling of on-shell closed string states to open superstring field theory. In particular we are going to discuss  the vacuum shift induced by a  physical closed string deformation and the corresponding open-closed effective couplings in RNS open superstring field theory \cite{Konopka:2016grr, Erler:2016ybs, Kunitomo:2015usa}. Focusing for simplicity on the open string NS sector, we have two main frameworks to discuss superstring field theories depending on whether we choose the dynamical string field to live in the large \cite{Berkovits, Berkovits:2004xh} or in the small Hilbert space \cite{Erler:2013xta}. 
 
We start our analysis with the WZW-like theory in the large Hilbert space \cite{Berkovits, Berkovits:2004xh}. This theory has the advantage of having 
microscopic vertices which are relatively simple, with no insertions of picture-changing operators. We then add a gauge-invariant open-closed term consisting of a simple vertex coupling an on-shell closed string with an off-shell open string, controlled by a deformation parameter $\mu$. The form which is more suited for studying the effective action is the so-called $t $-Ellwood invariant which couples a physical closed string in the (total) --1 picture in the small Hilbert space with an off-shell dynamical open string in the large Hilbert space and has been first discussed by Michishita \cite{Michishita:2004rx}. Just as in the bosonic string case \cite{OC-I} this term is a tree-level tadpole which destabilizes all the  vacua  of the theory, which have to be shifted, in order to cancel the tadpole. We show that upon expanding around a given vacuum shift solution (if such a solution exists) the theory remains structurally the same as the undeformed one but it is characterized by a new deformed BRST charge $Q_\mu$ which anticommutes with $\eta$ when the vacuum shift equations are satisfied. 

Then we analyze the structure of the vacuum shift equations perturbatively in the deformation parameter $\mu$ and we attempt to solve them by fixing the standard gauge $b_0=\xi_0=0$ outside of the kernel of $L_0$. In doing so we remain with equations in Ker$(L_0)$ which, order by order in $\mu$, are obstructions to the existence of the full solution, analogously to what happens for boundary marginal deformations  \cite{Vosmera:2019mzw, Maccaferri:2019ogq, Mattiello:2019gxc}. 
We derive a set of sufficient conditions on the closed string deformation which ensure that such obstructions are vanishing and a vacuum shift solution exists. These conditions are nicely interpreted as vanishing conditions for amplitudes involving an arbitrary number of deforming closed strings and a single physical open string. As it turns out, these amplitudes are naturally written in the large Hilbert space and (denoting $P_0$ the projector on the kernel of $L_0$) they have an interesting  structure involving symmetric insertions of ``dual'' homotopy operators $h$ and $\tilde h$  associated to the mutually commuting derivations $Q$ and $\eta$
\be
    h&=&\frac{b_0}{L_0}(1-P_0)\\
\tilde h&=&\frac{\tilde b_0}{L_0}(1-P_0)=\frac{\xi_0L_0-X_0b_0}{L_0}(1-P_0)\\
\,[Q,h]&=&[\eta,\tilde h]=1-P_0,
\ee
where (just like $b_0=[\eta, \xi_0b_0]$ with $[Q,b_0]=L_0$) we have defined  $\tilde b_0\equiv [Q,b_0\xi_0]$ with $[\eta, \tilde b_0]=L_0.$ This $\eta$-$Q$ symmetry is a consequence of the ${\cal N}=2$  structure which is at the heart of the WZW theory \cite{Berkovits:1994vy, Berkovits} and in fact something very analogous to the homotopy operators $h$ and $\tilde h$ have been already discussed  in  \cite{Kroyter:2012ni}, in the study of the BV quantization of the free WZW theory, where instead of $\xi_0 b_0$ the zero mode of $\xi b(z)$ was used.
In our approach, where we  solve perturbatively the equation of motion,  these operators come  into play rather naturally in the emerging perturbation theory even using the standard gauge fixing $\xi_0=b_0=0$, provided we remain outside the kernel of $L_0$ in internal propagators, as we should. After having discussed the existence of a (perturbative) vacuum shift solution, we turn our interest to the construction of the effective action for the modes in the kernel of $L_0$ (the ``physical '' or --at zero momentum-- the ``massless'' fields).   To build the effective action we  follow  \cite{Erbin:2019spp,  Maccaferri:2018vwo} and, order by order in perturbation theory, we obtain the effective couplings between the deforming closed string state and the massless open strings. The obtained couplings also include an effective tadpole which couples a single massless open string to several deforming closed strings and we consistently find out that this tadpole is precisely given by the above-mentioned obstructions to the vacuum shift, precisely following the same logic as in the bosonic string \cite{OC-I}. A rather sharp difference however, compared to the bosonic case, is that the presence of the two dual propagators $h$ and $\tilde h$ and the elementary multi-string vertices
make the perturbation theory expanding extremely fast and it becomes too cumbersome to go  higher than the first few orders. Nevertheless  we find out that, just as it was happening for the obstructions, this perturbation theory is still naturally organized in a way which is completely symmetric in $\eta$ and $Q$ and the associated homotopies $\tilde h$ and $h$, hinting to an all-order structure which we are however unable to  characterize.
 
Because of this, differently from the bosonic case \cite{OC-I}, we don't  give all-order results for the full open-closed effective action as we did in general for theories which are based on a cyclic $A_\infty$ structure \cite{Erbin:2020eyc}, see also\cite{Koyama:2020qfb}. Therefore, before attacking the explicit evaluation of the first non-trivial open-closed couplings (which are discussed in the third paper\cite{OC-III}) we address the same problem in the $A_\infty$ open superstring field theory in the small Hilbert space \cite{Erler:2013xta}, which can be obtained from the WZW theory in the large Hilbert space by  gauge fixing the $\eta$ gauge invariance and performing a field redefinition \cite{Erler:2015rra, Erler:2015uoa, Erler:2015uba}.
 
 On the world-sheet, the $A_\infty$ theory is more complicated compared to the WZW one, essentially because its multi-string vertices are constructed iteratively using integrated non-local insertions of the picture-changing operator $X=[Q,\xi]$ and its primitive $\xi$. However the great advantage of the $A_\infty$ theory is its build-in cyclic homotopy structure which allows to switch-on the co-algebra language and to handle at once  the whole tree-level perturbation theory.  Quite surprisingly no one ever described the analog of the Ellwood invariant \cite{Hashimoto:2001sm, Gaiotto:2001ji, Ellwood:2008jh} for the $A_\infty$ theory. In fact, despite the theory is still based on Witten star product it turns out that the usual insertion of a physical closed string at the midpoint of the identity string field does not commute with  the multi-string products, due to the non-local picture-changing insertions. However this failure of commutativity can be ``corrected'' by adding a whole tower of new products $E_k\left(\Psi^{\otimes k}\right)$ coupling the  physical closed string to multiple open strings. These new products $E_k$ can be upgraded to coderivations  ${\bf E}_k$ in the tensor algebra and, just  as it happens for the Ellwood invariant in Witten bosonic OSFT \cite{Erbin:2020eyc, Koyama:2020qfb},  it turns out that their sum ${\bf E}=\sum_{k=0}{\bf E}_k$ is nilpotent ${\bf E}^2=0$ and it commutes with the total nilpotent coderivation ${\bf M}$ describing the fundamental open string products of the theory. This new observable belongs to the general class of observables that have been discussed in \cite{Erbin:2020eyc} (and further properties will be discussed in \cite{Schnabl:2020xx}) for generic $A_\infty$ theories. Moreover, just as the open string products ${\bf M}$ can all be encoded in a (large-Hilbert space) cohomomorphism $\bf G$ acting on the coderivation of the free theory ${\bf Q}$ as $
{\bf M}={\bf G}^{-1}{\bf Q}{\bf G},
$\cite{Erler:2013xta}
the ``Ellwood'' coderivation ${\bf E}$ can be similarly obtained as
\be
{\bf E}={\bf G}^{-1}{\bf e}{\bf G},
\ee
where ${\bf e}$ is the coderivation associated to the simple 0-string product given by the same midpoint insertion which we used in the WZW theory. This has the consequence that, just as the WZW  and the $A_\infty$ theory are related by a partial gauge fixing of the former and a field redefinition \cite{Erler:2015rra, Erler:2015uoa, Erler:2015uba}, the same is true for the WZW theory deformed by $e$ and  the $A_\infty$ theory deformed by $\bf E$. 
To compute the first few terms of the effective action one can simply follow \cite{Maccaferri:2019ogq} but in fact it is much more efficient working at the level of co-algebras where we can straightforwardly apply the {\it vertical decomposition} for strong deformation retracts discussed in detail in  \cite{Erbin:2020eyc}  to write down in closed form the complete infinite tower of effective open-closed (tree-level) couplings that one obtains by integrating out the open string field outside the kernel of $L_0$.
 
The paper is organized as follows. In section \ref{sec:3} we repeat what we did for the bosonic string in \cite{OC-I}  in the superstring context using the NS WZW theory in the large Hilbert space, deforming it with the open-closed invariant. We first discuss the tadpole removal and then we construct perturbatively  the open-closed couplings.  We conclude section \ref{sec:3} with an explicit computation of a mass deformation induced by a change in the compactification radius for a non-BPS d1 brane. This parallels the corresponding computation for the bosonic string which we presented in \cite{OC-I}. In section \ref{sec:MunichObs} we  construct the microscopic open-closed coupling ${\bf E}$ corresponding to the Ellwood invariant in the $A_\infty$ theory in the small Hilbert space.  After having constructed the observable ${\bf E}$ we show that this indeed coincides with the  observable that one gets by partially gauge fixing the WZW theory with the previously discussed open closed coupling \cite{Michishita:2004rx} and performing the field redefinition described in \cite{Erler:2015rra, Erler:2015uoa, Erler:2015uba}.  In section \ref{sec:5} we systematically write down the open-closed couplings in the $A_\infty$ theory to all orders taking advantage of the co-algebra description. We end up in section   \ref{sec:7} with some discussion on future extensions of the presented material.

The relation between the open-closed couplings of the WZW and the $A_\infty$ theory, together with ${\cal N}=2$ localization techniques will be the main theme of the third and final paper of this series \cite{OC-III}.
 
\section{Open-closed effective couplings in WZW-like open superstring field theory}\label{sec:3}
In this section we are going to perform an  analysis of the bulk deformation in the case of the WZW superstring theory in the NS sector, analogously to what we did in the bosonic case \cite{OC-I}.
\subsection{Tadpole shift}
The WZW theory coupled to the superstring version of the Ellwood-invariant in the large Hilbert space \cite{ Ellwood:2008jh, Michishita:2004rx}\footnote{This is the  ``$t$-form'' of the Ellwood invariant. The equivalent  $\eta$ and $Q$ versions \cite{Erler:2013wda} are less convenient for our work.} has the following action
\be
S^{(\mu)}[\Phi]=-\int_0^1dt\aver{A_t,\eta A_Q-\mu e},\label{action-pack}
\ee
where we have defined the Maurer-Cartan forms
\be
A_Q&=&e^{-t\Phi}Qe^{t\Phi}\\
A_t&=&e^{-t\Phi}\del_te^{t\Phi}=\Phi
\ee
and $\Phi$ is the picture and ghost number zero open string field in the large Hilbert space and in the NS sector.
$Q$ and $\eta$ are respectively the zero modes of the BRST current  and the $\eta$ ghost of a given RNS superstring background and they are mutually commuting nilpotent derivations
\be
Q^2=0,\quad[\eta,Q]=0,\quad \eta^2=0.
\ee
Moreover
$$e\equiv V(i,-i) I $$
is the identity string field $I$ with a midpoint insertion of an $h=(0,0)$ primary $V(z,\bar z)$ at total picture $-1$, which obeys
\be
\eta e&=&0\\
Q e&=&0.
\ee
These properties guarantee that the operator $\aver{\Phi,e}$ is gauge invariant under the infinitesimal gauge transformations
\be
\delta e^\Phi&=&Q\Lambda_1 e^\Phi+e^\Phi \eta\Lambda_2\\
&\updownarrow&\0\\
\delta\Phi&=&\frac{\ad_\Phi}{e^{\ad_\Phi}-1}Q\Lambda_1+\frac{\ad_\Phi}{1-e^{-\ad_\Phi}}\eta \Lambda_2,
\ee
where $\ad_\Phi \chi\equiv[\Phi,\chi]$.
The action \eqref{action-pack} can be explicitly rewritten in terms of infinite interaction vertices for the open string field $\Phi$ 
\begin{align}
S^{(\mu)}[\Phi] =\frac{1}{2}\aver{\eta\Phi, Q\Phi}+{\cal I}(\Phi)+\mu\aver{\Phi, e}\,,
\label{eq:actRHS}
\end{align}
where the interacting part of the action is explicitly given by
\be
{\cal I}(\Phi)=\sum_{k=1}^\infty \frac{(-1)^k}{(k+2)!}\aver{\eta\Phi,\,\ad_{\Phi}^kQ\Phi}
\ee 
and we have denoted $$\ad_\Phi\chi\equiv[\Phi,\chi].$$
Since the action has been deformed by a gauge invariant operator, the gauge symmetry of the action is unchanged, but the equation of motion acquires a source term
\be
\eta Q\Phi+{\cal J}(\Phi)=\mu e\label{mu-eom-wzw}.
\ee
The interacting part of the equation of motion can be compactly written starting with the formal power series expansion
\be
G(z)\equiv\sum_{n=1}^\infty\frac{z^n}{(n+1)!}=\frac{e^z-1}{z}-1
\ee
as
\be
{\cal J}(\Phi)&=&\left[G\left(\ad_\Phi\right)\eta G\left(-\ad_\Phi\right)+G\left(\ad_\Phi\right)\eta+\eta G\left(-\ad_\Phi\right)\right]Q\Phi\\
&=&-\frac12[\eta\Phi,Q\Phi]+\frac16\left([\eta\Phi,[\Phi,Q\Phi]]-\frac12[\Phi,[\eta\Phi,Q\Phi]]+\frac12[\Phi,[\Phi,\eta Q\Phi]]\right)+O(\Phi^4)\0.
\ee
Just as in the bosonic case, because of the tadpole $\mu e$, $\Phi_0=0$ is not a solution anymore, so if we want to study the physics that is induced by the $\mu$-deformation we have  to shift the vacuum  to a new equilibrium point. Let's first address this problem at a formal level. To this end  we rewrite the EOM \eqref{mu-eom-wzw} in the more standard WZW form (see for example \cite{Maccaferri:2018vwo})
\be
\frac{e^{\ad_\Phi}-1}{\ad_\Phi}\left(\eta(e^{-\Phi}Qe^{\Phi})-\mu e\right)=0,\quad\leftrightarrow\quad\eta(e^{-\Phi}Qe^{\Phi})-\mu e=0,\label{pack-eom}
\ee
where we have used that the operator 
\be
\frac{e^{\ad_\Phi}-1}{\ad_\Phi}=1+\sum_{n=1}^\infty \frac{(\ad_\Phi)^n}{(n+1)!}\0
\ee
is invertible (notice also that by construction $\ad_\Phi e=0$).
Assume now we have found a solution $\Phi_\mu$ to \eqref{pack-eom}. By writing the group element $e^\Phi$ as \cite{Erler:2013wda}
\be
e^\Phi=e^{\Phi_\mu}e^{\tilde\phi}\quad&\leftrightarrow&\quad\Phi=\Phi_\mu+\phi\\
\tilde\phi=\log\left(e^{-\Phi_\mu}e^{\Phi_\mu+\phi}\right)\quad&\leftrightarrow&\quad\phi=\log\left(e^{\tilde\phi}e^{\Phi_\mu}\right)-\Phi_\mu
\ee
the action becomes \cite{jakub-thesis}
\be
S^{(\mu)}[\Phi]&=&S^{(\mu)}[\Phi_\mu]-\aver{\tilde\phi,\eta\left(e^{-\Phi_\mu}Qe^{\Phi_\mu}\right)-\mu e}-\int_0^1dt\aver{\tilde A_t,\eta\tilde A_{Q_\mu}}\0\\
&=&S^{(\mu)}[\Phi_\mu]-\int_0^1dt\aver{\tilde A_t,\eta\tilde A_{Q_\mu}},
\ee
where the equation of motion has been used to set the tadpole to zero and we have defined
\be
\tilde A_t&=&e^{-t\tilde\phi}\del_te^{t\tilde\phi}=\tilde\phi\\
\tilde A_{Q_\mu}&=&e^{-t\tilde\phi}Q_\mu e^{t\tilde\phi}\\
Q_\mu&=&Q+\ad_{e^{-\Phi_\mu}Qe^{\Phi_\mu}}.
\ee
Notice that the shifted kinetic operator $Q_\mu$ is nilpotent even without assuming that $\Phi_\mu$ solves the tadpole-sourced equation of motion. This may look odd but in fact the need for a proper solution is contained in 
\be
[Q_\mu,\eta]=\ad_{\eta\left(e^{-\Phi_\mu}Qe^{\Phi_\mu}\right)}=\ad_{\mu e}=0,
\ee
which parallels the analogous bosonic condition  discussed in \cite{OC-I}. In other words, when we expand around a proper vacuum shift $\Phi_\mu$, then  $Q_\mu$ is a nilpotent operator which maps the small-Hilbert space into itself (despite the fact that $e^{-\Phi_\mu}Qe^{\Phi_\mu}$ itself is not in the small-Hilbert space). So the shifted theory is characterized by the pair of mutually commuting nilpotent operators $Q_\mu$ and $\eta$ 
\be
Q_\mu^2=0,\quad[Q_\mu,\eta]=0,\quad\eta^2=0
\ee
and has thus the same algebraic structure as the initial theory (without the closed string deformation) but a different BRST charge. 
Therefore the situation is effectively the same as in the simpler bosonic case \cite{OC-I}.

Now, as in the bosonic case, we can search for the vacuum shift  $\Phi_\mu$  perturbatively in $\mu$
\be
\Phi_\mu=\sum_{\alpha=1}^\infty \mu^\alpha \phi_\alpha.
\ee
The vacuum shift equations \eqref{mu-eom-wzw}  reads
\be
\eta Q\phi_1&=& e\\
\eta Q\phi_2&=&\frac12[\eta\phi_1,Q\phi_1]\\
\eta Q\phi_3&=&\frac12[\eta\phi_1,Q\phi_2]+\frac12[\eta\phi_2,Q\phi_1]\0\\
&&-\frac16\left([\eta\phi_1,[\phi_1,Q\phi_1]]-\frac12[\phi_1,[\eta\phi_1,Q\phi_1]]+\frac12[\phi_1,[\phi_1,\eta Q\phi_1]]\right)\\
&\cdots&\0,
\ee
and we can try to solve them iteratively. Again the first equation is not always solvable. A possible solution can be searched for by  fixing the gauge\footnote{Since $L_0\neq 0$ by the presence of $(1-P_0)$ setting $h=0$ is the same as setting $b_0(1-P_0)=0$, i.e. Siegel gauge outside the kernel of $L_0$. This is a partial gauge fixing, because it does not touch the kernel of $L_0$ which remains unfixed.} $h\Phi_\mu=\xi\Phi_\mu=0$ where
\be
h&=&\frac{b_0}{L_0}(1-P_0)\\
\xi&\equiv&\xi_0(1-P_0)
\ee
 in the form
\be
\phi_1=-\xi h e+\varphi_1,
\ee
where $P_0\varphi_1=\varphi_1$.
By acting with $\eta Q$ we find that the condition for the existence of a solution (to this order) is
\be
P_0(\eta Q \varphi_1-e)=\eta Q \varphi_1-P_0e=0.
\ee
As in the bosonic case, a sufficient condition (which is also necessary in the zero momentum sector of the open string) is to set
\be
P_0e=0.
\ee
 Going further, if we insist in finding solutions which do not excite the kernel of $L_0$, at second order in  $\mu$ we get the equation
\be
\eta Q\phi_2=-\frac12[he,\tilde h e],
\ee
where we have found convenient to define  
\be
\tilde h\equiv[Q,h\xi]=\xi-X_0 h=\left(\xi_0-X_0\frac{b_0}{L_0}\right)(1-P_0),
\ee
being $X_0$  the zero mode of the picture raising operator $X_0=[Q,\xi_0]$.
The operator $\tilde h$ obeys
\be
\,[Q,\tilde h]&=&0\\
\,[\eta,\tilde h]&=&1-P_0\\
\tilde h h&=&\xi h=\xi_0\frac{b_0}{L_0}(1-P_0).\label{diff}
\ee
In particular we have
\begin{align}
Q \,\tilde h h e\,&=-\tilde h e\\
\eta\, \tilde h h e\,&= h e.
\end{align}

It is suggestive to realize that $\tilde h$ is a propagator in the ``dual'' small Hilbert space (which is identified with the kernel of $Q$, rather than $\eta$), where $\eta$ (instead of $Q$)  is the kinetic operator. Indeed just as  we have $b_0=[\eta, \xi_0 b_0]$ with $[Q,b_0]=L_0$, in the kernel of $Q$ we can define \footnote{We could have defined, as in \cite{Kroyter:2012ni}, $\tilde b_0'\equiv d_0\equiv[Q, (b\xi)_0]$ where $(b\xi )_0$ is the zero mode of the conformal field $b\xi (z)$ and get analogous properties. Then we would have had a slightly different  $\eta$-homotopy $\tilde h'\equiv \frac{d_0}{L_0}(1-P_0)$, resulting in a  total propagator different from \eqref{diff}   $\tilde h' h\neq \xi_0\frac{b_0}{L_0}(1-P_0)$.  In other words we are still fixing the standard  $\xi_0=b_0=0$  gauge (outside the kernel of $L_0$) but we are anyhow able to display a complete $\eta-Q$ symmetry of the amplitudes. }
\be
\tilde b_0=[Q,b_0\xi_0]=\xi_0L_0-X_0b_0
\ee
 which gives 
 \be
 [\eta,\tilde b_0]=L_0
 \ee
  so that 
\be
\tilde h=\frac{\tilde b_0}{L_0}(1-P_0)=\xi-X_0 h.
\ee
Another interesting relation in this regard is $L_0=[Q,[\eta,\xi_0 b_0]]=[\eta,[Q,b_0 \xi_0]]$.
Then it is not difficult to see that we have a solution
\be
\phi_2=\frac12\xi h[he,\tilde h e]=\frac12\tilde h h[he,\tilde h e],
\ee
provided the second order obstruction vanishes
\be
P_0[he,\tilde h e]=0.
\ee
Continuing perturbatively (and systematically setting to zero the possible massless corrections) we get an infinite set of sufficient conditions for the existence of the vacuum shift for the closed string which we summarize here for convenience
\begin{align}
&0=P_0 e\0\\
&0=P_0[he,\tilde h e]\label{obstr-wzw}\\
&0 =P_0\left(\frac14[he,\tilde h[he,\tilde h e]]+\frac14[\tilde h e,h[he,\tilde h e]]-\frac16[he,[\tilde h h e,\tilde h e]]+\frac1{12}[\tilde h h e,[he,\tilde h e]]\right)\0\\
&\quad\vdots\0
\end{align}
As in the bosonic case these conditions should capture the vanishing of S-matrix elements between the deforming bulk closed strings and a single massless open string. Notice that these amplitudes are written explicitly in the large Hilbert space and make use of both propagators $h$ and $\tilde h$ in a completely symmetric way. It would be interesting (but beyond the scope of this paper) to systematically investigate how these large Hilbert space amplitudes relate to the more familiar ones in the small Hilbert space.

To conclude this subsection, before discussing the effective action approach, we would like to spend  few words on the Ellwood invariant we use and on the projector condition $P_0 e=0$. To this end  let's start considering a NS-NS closed string deformation.
This can be constructed starting from a set of (holomorphic) $h=1/2$ superconformal primaries $\mathbb{U}^a_\frac{1}{2}$ and their anti-holomorphic mirrors $\bar{\mathbb{U}}^b$ by considering the picture $-1$ closed string state\footnote{The subscript ``NS'' stands for ``NS-NS''.}
\be V_{\rm NS}(z,\bar z)=V_{ab}\left(\eta \mathcal{U}^a(z)\bar Q \mathcal{\bar U}^b(\bar z)+Q \mathcal{U}^a(z)\bar \eta  \mathcal{\bar U}^b(\bar z)\right)
\ee
where we have defined 
\be
\mathcal{U}^a(z)\equiv\xi c\mathbb{U}^a_\frac{1}{2} e^{-\phi}(z)=c\gamma^{-1}\mathbb{U}^a_\frac{1}{2}(z) ,
\ee
and similarly for $\mathcal{\bar U}^b$.
Assuming a generic gluing condition $\bar{\mathbb{U}}^b_\frac{1}{2}\to \Omega^b_{\,\,\,c}\,\mathbb{U}^c_\frac{1}{2}$ the bulk field $V(z,\bar z)$ turns into the bilocal chiral field 
\be
\tilde V_{\rm NS}(z,z^*)=\tilde V_{ab}\left(\eta \mathcal{U}^a(z)\, Q \mathcal{U}^b(z^*)+Q \mathcal{U}^a(z) \,\eta  \mathcal{U}^b( z^*)\right),\label{eq:eNSNS}
\ee
where now the polarization also includes the gluing matrix
\be
\tilde V_{ab}=V_{ac}\Omega^c_{\,\,\,b}.
\ee
Computing the $\eta$ and $Q$ variations we  find the standard relations
\be
\eta U&=&c \mathbb{U}_{\frac12}\,e^{-\phi}=c \mathbb{U}_{\frac12}\,\delta(\gamma)\\
Q U&=&c \mathbb{U}_{1}-e^\phi\eta   \mathbb{U}_{\frac12}=c \mathbb{U}_{1}-\gamma\mathbb{U}_{\frac12},
\ee
where $\mathbb{U}_1$ is the worldsheet matter superpartner of $\mathbb{U}_{\frac12}$ 
\be
T_F(z)\mathbb{U}_{\frac12}(w)=\frac{1}{z-w}\mathbb{U}_1(w)+{\rm regular}.
\ee
The open string field $e$ is then given by
\be
e=U_1^\dagger\, \tilde V(i,-i)\ket0,\label{e-ell}
\ee
where the (twist-invariant) operator $U_1^\dagger$ is given by 
\be
U_1^\dagger=\exp\left(\sum_{n\geq 1}v_nL_{-2n}\right),\label{Udagger}
\ee
where $v_n$ are known (but unimportant in our analysis) coefficients \cite{wedges}.
The Fock space state $P_0 e$ can be explicitly computed as in the bosonic case and it gives
\be
P_0 e=\tilde V_{ab}\, c\del c\,\mathbb{W}_{\frac12}^{ab}\,e^{-\phi}(0)\ket0,
\ee
where the NS-boundary field $\mathbb{W}$ is given by the weight 1/2 component of the bulk-boundary OPE 
\be
\mathbb{W}_{\frac12}^{ab}\left(x\right)=\lim_{z\to z^*}(z-z^*)\left(\mathbb{U}^a_{\frac12}(z)\mathbb{U}_{1}^{b}(z^*)-\mathbb{U}^a_{1}(z)\mathbb{U}_{\frac12}^{b}(z^*)\right), \quad x={\rm Re}\,z.
\ee
Of course it is possible that, for a given bulk field $V(z,\bar z)$ and a given gluing condition $\Omega$ the boundary field $\mathbb{W}_{\frac12}$ is vanishing, and in this case the vacuum shift is unobstructed (to first order).
In case of a RR deformation we can use the picture $(-1/2,-1/2)$
\be
V_{\rm RR}(z,\bar z)=V_{\alpha\bar\beta}\, cS^\alpha e^{-\phi/2}(z)\, \bar c \bar S^{\bar\beta} e^{-\bar \phi/2}(\bar z),
\ee
where $\alpha$ and $\bar\beta$ are the spinor indices in a given chirality dictated by the chosen GSO projection.
The corresponding holomorphic bilocal field will depend on the gluing condition for the spin field  which we will generically take to be $\bar S^{\bar\beta}(\bar z)\to F^{\bar \beta}_{\, \,\,\tilde \beta} S^{\tilde \beta}(z^*)$, where the chirality of $\tilde \beta$ depends on the boundary conditions
\be
\tilde V_{\rm RR}(z,z^*)=\tilde V_{\alpha\tilde\beta}\, cS^\alpha e^{-\phi/2}(z)\, cS^{\tilde\beta} e^{-\phi/2}(z^*),\label{eq:eRR}
\ee
where, analogously to the NS-NS case
\be
\tilde V_{\alpha\tilde\beta}=V_{\alpha\bar\beta} F^{\bar \beta}_{\, \,\,\tilde \beta}\,\, .
\ee
In this case we easily find that (assuming zero momentum in the open string sector) the only possible outcome is
\be
P_0 e=\tilde V_{\alpha\tilde\beta}\,c\del c\, \mathbb{W}^{\alpha\beta}_{\frac12}\, e^{-\phi}(0) \ket 0,
\ee
where the NS open string field $\mathbb{W}^{\alpha\beta}_{\frac12}$ is given by the $h=1/2$ contribution of the R-R bulk-boundary OPE
\be
\mathbb{W}_{\frac12}^{\alpha\tilde\beta}e^{-\phi}\left(x\right)=\lim_{z\to z^*}(z-z^*)\left(S^\alpha e^{-\frac{\phi}{2}}(z)S^{\tilde\beta} e^{-\frac{\phi}{2}}(z^*)\right),\quad x={\rm Re}\,z.
\ee
So we see that in both cases of NS-NS or R-R deformations, the first order obstruction to the vacuum shift is associated to the creation of a physical NS open string field via the bulk-boundary OPE.

\subsection{Effective action and open-closed couplings in the WZW theory}
We now consider integrating out the states outside the kernel of $L_0$ from the deformed action \eqref{action-pack}.
So we write
\be
\Phi=P_0\Phi+(1-P_0)\Phi\equiv\varphi+R.
\ee
The equation of motion for $R$ is given by
\be
\eta Q R=-{\cal J}(\varphi+R)+\mu e\equiv-{\cal J}_\mu(\varphi+R).\label{massive-eom-mu}
\ee
Now we fix the gauge 
\be
h R=\tilde h R=0,\quad\leftrightarrow\quad b_0R=\xi_0R=0
\ee
and we act on the $R$-equation \eqref{massive-eom-mu} with the full propagator $\tilde h h$. In doing so we miss a part of the $R$ equation (the out of gauge equation) which will be however accounted for by the final effective equation for the massless field $\varphi$, see \cite{Erbin:2020eyc}. This gives the ``integral'' equation 
\be
R=\tilde h h\,{\cal J}_\mu(\varphi+R).
\ee
As in the bosonic string \cite{OC-I}, the solution $R=R_\mu(\varphi)$ to this equation is directly related to the corresponding solution $R_{\mu=0}(\varphi)$ in the undeformed $\mu=0$ theory. Indeed it is easy to see that we have
\be
\Phi_\mu(\varphi)=\varphi+R_\mu(\varphi)=\Phi_{\mu=0}(\varphi-\mu h\tilde h e).\label{subst}
\ee
Explicitly the first few terms of $R_{\mu}(\varphi)$ are given by
\begin{align}
R_\mu(\varphi)&=\sum_{k=0}^\infty \sum_{\alpha=0}^\infty\mu^\alpha\,R_{k ,\alpha}\\
R_{0,0}&=R_{1,0}=0\\
R_{0,1}&=-\tilde h he\\
R_{2,0}&=-\frac{\tilde h h}{2}[\eta\varphi,Q\varphi]\\
R_{1,1}&=-\frac{\tilde h h}{2}\left([\eta\varphi,\tilde h e]-[he,Q\varphi]   \right)\\
R_{0,2}&=\frac{\tilde h h}{2}[he,\tilde h e].\\
&\vdots&\0
\end{align}
Notice how $R_{k ,\alpha}$ is obtained from $R_{(k+\alpha),0}$ by doing $\alpha$ substitutions $\varphi\to -\tilde h h e$ in all possible ways, according to \eqref{subst}. Already looking at these low order terms, we can realize that a natural way to interpret the double summation in $\alpha$ (number of closed strings insertions) and $k$ (number of open string string insertion) is an expansion in  $2\alpha+k$ which is how the usual string perturbation theory is typically understood. Higher order terms can be obtained straightforwardly but the reader can check that the corresponding perturbation theory becomes cumbersome very soon.
It would be clearly desirable to have a closed form expression for the solution $R_{\mu=0}(\varphi)$ as in the case of theories based on a cyclic $A_\infty$ algebra (like the bosonic OSFT and the small Hilbert space theory which we focus on in the next sections). 
If we keep these first few orders (renaming $\varphi\to\lambda\varphi$ to explicitly have an open string counting parameter) we arrive at the final form of the effective action as
\be
S_{\rm eff}^{(\mu)}[\lambda\varphi]&=&\left(-\frac{\mu^2}{2}\aver{e,\tilde h h e}+O\left(\mu^3\right)\right)\0\\
&+&\lambda \left(\mu\aver{e,\varphi}-\frac{\mu^2}{2}\aver{[he,\tilde h e],\varphi}+O\left(\mu^3\right)\right)\0\\
&+&\lambda^2\left(\frac12\aver{\eta\varphi,Q\varphi}-\frac{\mu}{2}\aver{\eta\varphi,\left[\tilde h he,Q\varphi\right]}+O\left(\mu^2\right)\right)\0\\
&+&\lambda^3\left(-\frac16\aver{\eta\varphi,[\varphi,Q\varphi]}+O\left(\mu\right)\right)\0\\
&+&\lambda^4\left(\frac1{24}\aver{\eta\varphi,[\varphi,[\varphi,Q\varphi]]}-\frac18\aver{[\eta\varphi,Q\varphi],\tilde h h[\eta\varphi,Q\varphi]}+O(\mu)\right)\0\\
&+&O(\lambda^5).\label{eq:SeffWZW}
\ee
At $O(\lambda^0)$ we find the non-dynamical cosmological constant which consists of purely closed string scattering off the initial D-brane. Then at $O(\lambda)$ we find the ``massless'' tadpole, which is the same as the obstructions to the full vacuum shift \eqref{obstr-wzw}. At $O(\lambda^2)$ we find the kinetic term for the (initially) massless fields $\varphi$. The closed string deformation gives rise to possible mass-terms consisting (at leading order) of a (open)$^2$-(closed) amplitude. The non vanishing of this amplitude is in turn a first order obstruction to the existence of an open string marginal deformation triggered by a physical $\varphi$. At higher order the purely open string effective action already studied in \cite{Maccaferri:2018vwo} gets corrected by infinite closed string corrections according to the general picture we have already encountered in the bosonic case.  
Notice that although the recipe to get the above effective couplings is rather straightforward,  it is not easy to grasp an all-order structure, which instead was very clear in the bosonic string analysed in \cite{OC-I}. For this reason, as a step towards unveiling the all-order structure of the WZW effective action, we will analyze the same situation in the closely related $A_\infty$ open superstring in the small Hilbert space where, following the general structure presented in \cite{Erbin:2020eyc}, we will be able to give all-order results on the effective open-closed couplings. However before delving into this rather technical endevour we will give a concrete example of how the above effective couplings (in particular the one responsible for the mass deformation of the open string spectrum) can be easily computed by simple generalization of what we have done for the bosonic string in \cite{OC-I}.
\subsection{Example: mass correction for the radius deformation of a non-BPS D-string in Type IIA}
Before going on with more extended all-order constructions, which will be the topic of the following sections, let us give a simple example on how the above terms in the effective action can be explicitly computed, paralleling what we did in the bosonic case \cite{OC-I}. The setting we want to consider here is a non-BPS D1-brane wrapped on a circle of radius $R$ in Type IIA string theory. The BCFT of this system has been studied in \cite{Sen:2003zf} at the critical radius $R=\sqrt{2\alpha'}$ where the weight $1/2$ GSO$(-)$ matter states $e^{\pm i \sqrt{\frac2{\alpha'}}Y}$ (in canonical --1 picture )\footnote{In our conventions $Y\equiv Y_L$ stands for the chiral part of the $Y(z,\bar z)$ compact free boson.} generate an exact moduli space connecting the D-string to a $D0$-$\overline{D0}$ system at maximal distance on the circle. In our example we will start at generic radius $R$ and we will add a closed string deformation 
of the form 
\be
\mu e=\frac{\mu}{4\pi i}U_1^\dagger\left(\eta \mathcal{U}(i)Q \mathcal{U}(-i)+Q \mathcal{U}(i)\eta \mathcal{U}(-i)\right)\ket{0},
\ee
see \cite{OC-I} and \cite{wedges} for the definition of the world-sheet operator $U_1^\dagger$. For the radius deformation we have to consider 
\be
\mathcal{U}(z)=c\xi e^{-\phi}\psi_y (z)
\ee
and therefore
\be
\eta\mathcal{U}&=&c\psi_y e^{-\phi}(z),\\
Q \mathcal{U}&=&c j_y(z)-e^{\phi}\eta \psi_y(z),
\ee
where 
\be
j_y(z)=i\sqrt{\frac{2}{\alpha'}}\del Y(z)
\ee
is the world-sheet SUSY partner of  $\psi_y$ 
\be
T_F(z)\psi_y(0)=\frac1z j_y(0)+{\rm regular}.
\ee
This  deformation is the superstring analog of what has been studied in \cite{OC-I} and it describes a modification of the compactification radius $R\to R+\delta R$. The relation between $\delta R$ and $\mu$ will be derived shortly.
But first of all we would like to check that the massless open string field 
\be
\varphi_0=\phi_0 \,c\xi e^{-\phi}j_y \label{GSO+}
\ee
remains massless. To this end we compute the induced mass term from \eqref{eq:SeffWZW}
\be
\frac{\alpha'}2\phi_0 m_0^2 \phi_0&=&-\frac{\mu}{2}\aver{\eta\varphi,\left[\tilde h he,Q\varphi\right]}=\frac{\mu}{2}\aver{e,\tilde h h\left[\eta\varphi,Q\varphi\right]}\0\\
&=&\frac{\mu}{2}\aver{e{\Bigg |}\,\xi_0\frac{b_0}{L_0}(1-P_0)\,{\Bigg|}\left[\eta\varphi,Q\varphi\right]}.
\ee
Thanks to the condition $P_0e=0$ (which can be readily verified by direct OPE) the above quantity can be manipulated in complete analogy to what has been done in \cite{OC-I} for the corresponding bosonic string computation. This gives the result
\be
&&\frac{\alpha'}2\phi_0 m_0^2 \phi_0\0\\
&=&\frac{\mu}{4\pi i}\phi_0^2\int_0^1dt\aver{ \left(c\psi e^{-\phi}(i)cj(-i)+cj(i)c\psi e^{-\phi}(-i)\right){\Big(}c(t)+c(-t){\Big)}\left(\psi e^{-\phi}(t)j(-t)+j(t)\psi e^{-\phi}(-t)\0\right)}\0\\
&=&\frac{\mu}{4\pi i}\phi_0^2\int_0^1dt4i(1+t^2)\aver{ \left(\psi e^{-\phi}(i)j(-i)+j(i)\psi e^{-\phi}(-i)\right)\left(\psi e^{-\phi}(t)j(-t)+j(t)\psi e^{-\phi}(-t)\0\right)}\0\\
&=&\frac{2\mu}{\pi}\phi_0^2\int_0^1dt(1+t^2)\left(\frac1{(t-i)^4}+\frac1{(t+i)^4}\right)=0.
\ee
Therefore we find that the mode $\phi_0$ remains massless under the radius deformation. This is what we expect since this mode  describes the Wilson line deformation which  is exactly marginal for every radius. It is interesting to notice that here in the superstring the integral is finite without the need of a regularization at $t\sim 0$ as it was the case for the bosonic string \cite{OC-I}. The reason is that, thanks to the fact that the interacting open strings are at different pictures, there is no zero-momentum negative-weight field propagating in the amplitude.

Next, we would like to compute the mass correction for the GSO(--) fields 
\be
\varphi_{n}^{\pm}(k)=\phi_n^{\pm}(k)\, c\xi e^{-\phi}{\cal P}_n^{\pm}\otimes \sigma_1\label{GSO-},
\ee
where we have set
\be
{\cal P}_n^{\pm}=e^{\pm i \frac{n}{R\sqrt{\alpha'}}Y}e^{ik\cdot X}
\ee
and we have explicitly included an internal Chan-Paton factor  $\sigma_1$\cite{Berkovits:2000hf}.\footnote{A trivial $2\times2$ identity matrix is also understood in \eqref{GSO+}} A cocycle factor is understood to make ${\cal P}$ and $e^{-\phi}$ effectively grassmann-odd (like $\psi$) \cite{Sen:2003zf}.
 We have also considered the field at finite momentum $k$ in order to be able to put $\varphi^{\pm}$ on-shell for all values of $R$ and $n$ so that the SFT amplitude will be computable without using the Schwarz-Christoffel map as in \cite{OC-I} and in the previous example. The mass-shell condition is given by
\be
k^2=\frac1{2\alpha'}-\frac{n^2}{R^2}
\ee
and under this condition \eqref{GSO-} is a physical weight-zero field. Taking into account the multiplicative Chan-Paton for the derivations $\eta$ and $Q$ \cite{Berkovits:2000hf}
\be
\eta&\to&\hat\eta=\eta\otimes\sigma_3\\
Q&\to&\hat Q=Q\otimes\sigma_3
\ee
we have
\be
\hat\eta\,\varphi_{n}^{\pm}(k)&=&\phi_n^{\pm}(k)\, c{\cal P}_n^{\pm} e^{-\phi}\otimes (i \sigma_2),\\
\hat Q\,\varphi_{n}^{\pm}(k)&=&\phi_n^{\pm}(k)\left[c\left(\pm\sqrt{2\alpha'}\frac{n}{R} \psi_y+\sqrt{2\alpha'}(k\cdot\psi)\right){\cal P}_n^{\pm}-e^\phi\eta{\cal P}_n^{\pm}\right]\otimes(i\sigma_2)\0\\
&=&\phi_n^{\pm}(k)\left[c\left(\pm\sqrt{2\alpha'}\frac{n}{R} \psi_y\right){\cal P}_n^{\pm}\right]\otimes(i\sigma_2)+ (\textrm{non-contributing terms}).
\ee
We are interested in the term of the effective action \eqref{eq:SeffWZW}
\be
\alpha' \delta m_\pm^2\phi_n^+(k)\phi_n^-(k')\delta(k+k')=\frac\mu2\left(\aver{e,\tilde h h\left[\hat\eta\varphi_n^+,\hat Q\varphi_n^-\right]}+ (+\leftrightarrow-)\right).\label{masspm}
\ee
The correlator gives four terms which are all equal and we end up with
\begin{align}
&\left(\aver{e,\tilde h h\left[\hat\eta\varphi_n^+,\hat Q\varphi_n^-\right]}+ (+\leftrightarrow-)\right)=4\times\frac{1}{2\pi i}\aver{Q\mathcal{U}(i)\eta\mathcal{U}(-i),\xi_0\frac{b_0}{L_0}\left(\hat\eta\varphi_n^+(1),\hat Q\varphi_n^-(-1)\right)}\0\\
&=\frac{8\sqrt{2\alpha'}\frac nR}{\pi}\phi_n^+(k)\phi_n^-(k')\int_0^1dt\,(1+t^2)\aver{j_y(i)\psi_ye^{-\phi}(-i){\cal P}_n^{-}e^{-\phi}(t)\psi_y{\cal P}_n^{+}(-t)}\times \frac12\tr[(i\sigma_2)^2]=(*),
\end{align}
where we have evaluated the $bc$ correlator and we have isolated an overall minus sign from the internal Chan-Paton's factors (with an understood normalization of the Chan-Paton's trace). Computing the remaining correlator (keeping in mind that $e^{-\phi}$ and ${\cal P}^{\pm}$ are effectively Grassmann odd thanks to their implicit cocycle factors) we finally obtain
\begin{align}
(*)=\frac{16 \alpha'\frac {n^2}{R^2}}{\pi}\phi_n^+(k)\phi_n^-(k')\delta(k+k')\int_0^1dt\frac1{1+t^2}=4\alpha'\frac {n^2}{R^2}\phi_n^+(k)\phi_n^-(k')\delta(k+k').
\end{align}
Therefore, from \eqref{masspm}, we find
\be
\delta m_\pm^2=2\mu\frac {n^2}{R^2}.\label{mass-shift} 
\ee
Now we can relate the SFT deformation parameter $\mu$ to the change in the compactification  radius $\delta R$ by matching \eqref{mass-shift} with the analogous formula one gets from the KK spectrum
\be
m_\pm^2(R)=\frac {n^2}{R^2}-\frac1{2\alpha'},
\ee
from which we get
\be
\delta  m_\pm^2=\frac {n^2}{(R+\delta R)^2}-\frac {n^2}{R^2}=-2\frac {n^2}{R^2}\frac{\delta R}{R}+O(\delta R^2).
\ee
Therefore we have found 
\be
\mu=-\frac{\delta R}{R}+O(\delta R^2),
\ee
a result which is completely analogous to the corresponding computation in bosonic SFT of \cite{OC-I}. In particular we find that that at the critical radius $R=\sqrt{2\alpha'}$ for $n=1$ the massless modes  $e^{\pm i \sqrt{\frac2{\alpha'}}Y}$ become tachyonic as the radius increases and the marginal direction connecting the non-BPS D1 brane to the $D0$-$\bar{D0}$ pair is lifted as expected. Following our general picture the non-vanishing of the above mass-shift amplitude is an obstruction to the existence of a solution representing an open string marginal deformation.
In the third paper \cite{OC-III} we will confront ourselves with $O(\mu)$ mass terms for more complicated D-branes setting and we will take advantage of extra worldsheet structure (the ${\cal N}=2$ worldsheet supersymmetry) which will drastically simplify the computation of the involved four-point function (which in this case was elementary). 

\section{A new observable in $A_\infty$ open superstring field theory}
\label{sec:MunichObs}

While the WZW-like open superstring field theory in the large Hilbert space yields interaction vertices which are extremely economical, at any given order they do not seem to exhibit any recognizable algebraic structure. As we have already emphasized in the previous section, this fact prevents us from obtaining all-order results on the vacuum shift and the effective open-closed vertices using the WZW-like open SFT deformed by adding the Ellwood invariant. Fixing the $\eta$-gauge symmetry by setting $\xi_0 \Phi=0$, it was shown that the WZW-like equation of motion can be rewritten in terms of an $A_\infty$ structure which, however, is not cyclic, so that it is not manifested at the level of action. At the same time, the small Hilbert space theory obtained in this way is known to be related by an explicit field redefinition \cite{Erler:2015rra,Erler:2015uba} to the ``Munich'' open superstring field theory \cite{Erler:2013xta}, which displays a very elegant structure of interactions expressed in terms of a cyclic $A_\infty$ structure (although this comes at a cost of having to deal with a somewhat complicated distribution of PCO insertions in the expanded vertices). A brief review of this theory in the formalism of tensor coalgebras \cite{Gaberdiel:1997ia} and its relation to the WZW-like theory is presented in Appendix \ref{app:MunichReview}. The algebraic simplicity of the Munich theory then makes it possible to conveniently package the whole perturbation theory using the homological perturbation lemma \cite{Kajiura:2001ng,Kajiura:2003ax} and therefore to present closed-form expressions for effective vertices at arbitrary order. However, if we want to exploit this machinery also for the all-order calculation of the effective open-closed vertices and the vacuum shift in the presence of an exactly marginal closed-string background, a suitable analogue of the Ellwood invariant for the $A_\infty$ theory needs to be discussed first.

The aim of this section will be to present a construction of an observable for the cyclic $A_\infty$ open superstring field theory, which is based on the string field $e$ used in the WZW theory given in \eqref{e-ell}, which is a midpoint insertion of an on-shell weight $(0,0)$ picture $-1$ closed-string primary on the identity string field $I$.\footnote{We would like to thank Ted Erler for very useful discussions on this topic. This construction was originally presented in the talk \cite{jak-talk} by one of the authors of this paper.} This will be called the \emph{bosonic Ellwood state}. Contrary to the case of the bosonic cubic OSFT, we will however see that computing the BPZ product of $e$ with the dynamical string field does \emph{not} give rise to a gauge-invariant quantity (the Ellwood invariant) of the $A_\infty$ open superstring field theory. Taking a lesson from the construction of the superstring products $M_k$ of the Munich theory, we will then attempt to define higher products $E_k$ for $k\geqslant 1$ so as to build the corresponding observable order by order in $\Psi$.
Eventually, we will recognize a recurrent pattern whose validity we will then proceed to establish to all orders in $\Psi$. Since all $E_k$ will turn out to be linear in $e$, also the resulting observable (the ``dressed Ellwood invariant'') will be linear in $e$.
We will also show that our observable is related by the field redefinition of \cite{Erler:2015rra,Erler:2015uba} to the $t$-Ellwood invariant of the partially gauge-fixed Berkovits' WZW-like open SFT. In Section \ref{sec:5}, we will see that adding this observable to the action yields a new theory which formally exhibits weak $A_\infty$ structure.

\subsection{Zero- and one-string products}

The bosonic Ellwood state $e$, as described above, is known to satisfy the properties \cite{Hashimoto:2001sm,Gaiotto:2001ji,Ellwood:2008jh}
\begin{subequations}
	\begin{align}
	0&=Qe\,,\\
	0&=m_2(\Psi,e)+m_2(e,\Psi)\,,\label{eq:P2}
	\end{align}
\end{subequations}
where $m_2$ denotes the Witten star product.
It is then straightforward to see that the corresponding bosonic Ellwood invariant $\omega_\text{S}(\Psi,e)$ changes non-trivially under the gauge transformation 
\eqref{eq:gaugeT}
of the $A_\infty$ open SFT (here $\omega_\text{S}$ denotes the symplectic form given by the BPZ product in the small Hilbert space). Indeed, while the gauge variation of $\omega_\text{S}(\Psi,e)$ at $\mathcal{O}(\Psi^0)$ still vanishes, at first order in $\Psi$ the variation reads
\begin{subequations}
	\begin{align}
	\omega_\text{S}\big(\Lambda,{M}_2({e},\Psi)+{M}_2(\Psi,{e})\big)&=\frac{1}{3}\omega_\text{S}\big(\Lambda,m_2(X_0e,\Psi)+m_2(\Psi,X_0e)\big)\\[2mm]
	&\neq 0
	\end{align}
\end{subequations}
because $X_0e$ can no longer be regarded as a local midpoint insertion of a closed string operator on the identity string field. Nevertheless, it is possible to quickly verify that this $\mathcal{O}(\Psi)$ anomaly can be corrected by replacing
\begin{align}
\omega_\text{S}(\Psi,e) \longrightarrow \omega_\text{S}(\Psi,e)+\frac{1}{2}\omega_\text{S}(\Psi,E_1(\Psi))\,,
\end{align}
where we define a new cyclic 1-product
\begin{align}
E_1(\Psi) &= -\mu_2(e,\Psi)-\mu_2(\Psi,e)\label{eq:E1}
\end{align}
in terms of the gauge 2-product $\mu_2$ (see \eqref{eq:mudef} for an explicit expression for $\mu_2$).
It is also an immediate consequence of the property \eqref{eq:P2} (and the fact that $e\in\mathcal{H}_\text{S}$) that the product $E_1$ is in the small Hilbert space. We have therefore managed to dress the bosonic Ellwood invariant by adding a $\mathcal{O}(\Psi^{\otimes 2})$ term defined in terms of a small Hilbert space cyclic 1-product $E_1$ so as to restore gauge invariance to the linear order in $\Psi$. Hence, it is reasonable to expect that by adding higher order terms defined using some $k$-products $E_k$ for $k\geq 2$, we can achieve restoration of gauge invariance to all orders in $\Psi$.

\subsection{Higher products}

In order to avoid lengthy expressions, we will find it more convenient to work using the tensor coalgebra notation introduced in Appendix \ref{app:coalg}. Denoting the desired quantity (which is to be gauge-invariant in the $A_\infty$ open SFT) by $\mathcal{E}(\Psi)$, we will see that it can be defined in terms of an odd cyclic small Hilbert space coderivation $\mathbf{E}$ (see Appendix \ref{app:coalg} for our conventions on the coalgebra machinery) as
\begin{subequations}
	\label{eq:obsMunich}
	\begin{align}
	\mathcal{E}(\Psi)&\equiv \sum_{k=0}^\infty\frac{1}{k+1}\omega_\text{S}(\Psi,E_k(\Psi^{\otimes k}))\\
	&= \int_0^1 dt\,\langle\omega_\text{S} |\, \pi_1\bs{\p}_t \frac{1}{1-\Psi(t)}\otimes \pi_1 \mathbf{E} \frac{1}{1-\Psi(t)}\,,
	\end{align}
\end{subequations}
where the ghost-number $2-k$, picture-number $k-1$ small Hilbert space products $E_k: \mathcal{H}^{\otimes k}\longrightarrow \mathcal{H}$ can be extracted as $E_k = \pi_1 \mathbf{E}\pi_k$ with $E_0\equiv e$ and we have introduced an arbitrary interpolation $\Psi(t)$ for $0\leqslant t \leqslant 1$ where $\Psi(0)=0$ and $\Psi(1)=\Psi$.
It was shown in \cite{Erbin:2020eyc} that such $\mathcal{E}(\Psi)$ is gauge-invariant (up to terms which vanish on-shell) whenever $\mathbf{E}$ commutes as a coderivation with $\mathbf{M}$, that is, whenever we have 
\begin{align}
[\mathbf{E},\mathbf{M}]=0\label{eq:[E,M]}
\end{align}
(more details are to be given also in \cite{Schnabl:2020xx}). In terms of the coderivations $\mathbf{E}_k$ and $\mathbf{M}_k$ corresponding to the products $E_k$ and $M_k$ (obviously we then have $\mathbf{E}=\sum_{k=1}^\infty\mathbf{E}_k$), the condition \eqref{eq:[E,M]} can be rewritten as
\begin{align}
\sum_{l=0}^{k-1} [\mathbf{E}_l, \mathbf{M}_{k-l}]=0\label{eq:[E,M]k}
\end{align}
for all $k\geq 1$. 

Let us start our construction of $\mathbf{E}$ by considering
an odd coderivation $\mathbf{e}$ which corresponds to the Ellwood state $e$ understood as a 0-string product. The coderivation $\mathbf{e}$  satisfies
\begin{subequations}
	\label{eq:midBos}
	\begin{align}
	[\mathbf{Q},\mathbf{e}]&=0\,,\label{eq:mid1M}\\
	[\mathbf{m}_2,\mathbf{e}]&=0\,,\label{eq:mid2M}
	\end{align}
\end{subequations}
and the fact that $e\in\mathcal{H}_\mathrm{S}$ gives us also $[\bs{\eta},\mathbf{e}]=0$.
Setting $\mathbf{E}_0\equiv \mathbf{e}$, we are therefore able to satisfy the condition \eqref{eq:[E,M]k} for $k=1$, namely  
\begin{align}
[\mathbf{E}_0,\mathbf{M}_1]=0\,.
\end{align}
We also have $[\bs{\eta},\mathbf{E}_0]=0$, namely that $\mathbf{E}_0$ is a small Hilbert space coderivation. Moreover, $\mathbf{E}_0$ is (trivially) cyclic.
Next, recalling the form \eqref{eq:E1} of the 1-product $E_1$, we are led to put
\begin{align}
\mathbf{E}_1 \equiv [\mathbf{E}_0,\boldsymbol{\mu}_2]\,.\label{eq:defE1}
\end{align}
First, note that $\mathbf{E}_1$ is really in the small Hilbert space: recalling that $[\bs{\eta},\bs{\mu}_2]=\mathbf{m}_2$ and using that $[\boldsymbol{\eta},\mathbf{E}_0]=0=[\mathbf{E}_0,\mathbf{m}_2]$,
we can indeed verify that $	[\boldsymbol{\eta},\mathbf{E}_1] =0$.
The above definition \eqref{eq:defE1} also makes it manifest that $\mathbf{E}_1$ is cyclic, because the commutator of two cyclic coderivations is again a cyclic coderivation. Finally, using that $[\mathbf{M}_1,\boldsymbol{\mu}_2]=\mathbf{M}_2$ and $[\mathbf{M}_1,\mathbf{E}_0]=0$, we can verify the condition \eqref{eq:[E,M]k} for $k=2$, namely that 
\begin{align}
[\mathbf{E}_0,\mathbf{M}_2]+[\mathbf{E}_1,\mathbf{M}_1]=0\,.
\end{align}
Similarly, by using the properties of $\mathbf{E}_0$, $\mathbf{E}_1$, we can convince ourselves (see Appendix \ref{app:E2} for a detailed calculation) that by defining
\begin{align}
\mathbf{E}_2 \equiv \frac{1}{2}\Big([\mathbf{E}_1,\boldsymbol{\mu}_2]+[\mathbf{E}_0,\boldsymbol{\mu}_3]\Big)\,,
\end{align}
we obtain $[\bs{\eta},\mathbf{E}_2]=0$, as well as the condition \eqref{eq:[E,M]k} for $k=3$. Hence, we are led to conjecture that for general $k>0$, the coderivation $\mathbf{E}_k$ can be defined recursively as
\begin{align}
\mathbf{E}_k \equiv \frac{1}{k}\bigg([\mathbf{E}_0,\boldsymbol{\mu}_{k+1}]+[\mathbf{E}_1,\boldsymbol{\mu}_{k}]+\ldots+[\mathbf{E}_{k-1},\boldsymbol{\mu}_2]\bigg)\equiv\frac{1}{k}\sum_{l=0}^{k-1}[\mathbf{E}_l,\bs{\mu}_{k+1-l}]\,.\label{eq:conjE}
\end{align}
Note that the coderivations $\mathbf{E}_k$ carry picture number $k-1$. 

We will now prove that the recursion \eqref{eq:conjE} indeed gives $\mathbf{E}$ such that $[\mathbf{E},\mathbf{M}]=[\boldsymbol{\eta},\mathbf{E}]=0$. To this end, let us introduce the generating functions
\begin{subequations}
	\begin{align}
	\mathbf{E}(t) &\equiv \sum_{k=0}^\infty t^k \mathbf{E}_k\,,\label{eq:Egen}\\
	\bs{\mu}(t) &\equiv \sum_{k=0}^\infty t^k \bs{\mu}_{k+2}\,.
	\end{align}
\end{subequations}
It is then straightforward to compute (see Appendix \ref{app:proofdiff} for details) that the recursion \eqref{eq:conjE} implies the following differential equation
\begin{align}
\frac{\p}{\p t} \mathbf{E}(t)  &= [\mathbf{E}(t),\bs{\mu}(t)]\,,\label{eq:diffE1}
\end{align}
for $\mathbf{E}(t)$, where \eqref{eq:Egen} gives an initial condition $\mathbf{E}(0)=\mathbf{e}$, while we also have $\mathbf{E}(1)=\mathbf{E}$. Solving \eqref{eq:diffE1}, we therefore find that
\begin{align}
\mathbf{E}=\mathbf{G}^{-1}\mathbf{e}\mathbf{G}\,,\label{eq:GeG}
\end{align}
where the cyclic cohomomorphism $\mathbf{G}$ is defined in \eqref{eq:cohomG}.
Recalling the expression \eqref{eq:GQG} for $\mathbf{M}$, as well as the relation \eqref{eq:GetaG} and the conditions \eqref{eq:midBos} on the coderivation $\mathbf{e}$ derived from the bosonic Ellwood state, it is then straightforward to verify that \eqref{eq:GeG} indeed gives $[\mathbf{E},\mathbf{M}]=[\boldsymbol{\eta},\mathbf{E}]=0$. Moreover, $\mathbf{E}$ is manifestly cyclic because $\mathbf{G}$ is a cyclic cohomomorphism. It also follows from \eqref{eq:GeG} and from $[\mathbf{e},\mathbf{e}]=0$ (this is true for any coderivation derived from a 0-string product) that the coderivation $\mathbf{E}$ is formally\footnote{Ignoring any potential divergences coming from closed-string collisions at the midpoint.} nilpotent, namely that we have
\begin{align}
[\mathbf{E},\mathbf{E}]=0\,.\label{eq:nilp}
\end{align}
We can therefore conclude that, at least formally, the products $E_k$ satisfy the relations of a weak $A_\infty$ algebra, which commutes with the $A_\infty$ algebra of the products $M_k$ (in the sense that $[\mathbf{E},\mathbf{M}]=0$).

Finally, for the sake of concreteness, let us expand the expression \eqref{eq:GeG} for $\mathbf{E}$, the associated products $E_k$, as well as the resulting observable $\mathcal{E}(\Psi)$ explicitly in terms of ${e}$ and $\mu_k$ for a couple of lowest orders in $\Psi$. We obtain
\begingroup\allowdisplaybreaks
\begin{subequations}
	\begin{align}
	\mathbf{E}_0 &= \mathbf{e}\,,\\[2mm]
	\mathbf{E}_1 &= [\mathbf{e},\boldsymbol{\mu}_2]\,,\\[1mm]
	\mathbf{E}_2 
	&=\frac{1}{2}\Big([[\mathbf{e},\boldsymbol{\mu}_2],\boldsymbol{\mu}_2]+[\mathbf{e},\boldsymbol{\mu}_3]\Big)\,,\\[1mm]
	&\hspace{0.2cm}\vdots\nonumber
	\end{align}
\end{subequations}
\endgroup
which in turn yields
\begingroup\allowdisplaybreaks
\begin{subequations}
	\label{eq:explE}
	\begin{align}
	E_0 &= e\,,\\[1mm]
	E_1(\Psi) 
	&= -\mu_2(e,\Psi)-\mu_2(\Psi,e)\,,\label{eq:E1expl}\\
	E_2(\Psi,\Psi)&=-\frac{1}{2}\Big({\mu}_3(e, \Psi,\Psi)+{\mu}_3(\Psi, e,\Psi)+{\mu}_3( \Psi,\Psi, e)\Big)+\nonumber\\
	&\hspace{2cm}+\frac{1}{2}\Big(
	{\mu}_2({\mu}_2(e, \Psi),\Psi)
	-{\mu}_2(e,{\mu}_2(\Psi,\Psi))
	+\nonumber\\
	&\hspace{4cm}
	+{\mu}_2({\mu}_2(\Psi, e) ,\Psi)
	+{\mu}_2(\Psi,{\mu}_2(e ,\Psi))
	+\nonumber\\
	&\hspace{4cm}
	-{\mu}_2({\mu}_2(\Psi,\Psi),e)
	+{\mu}_2(\Psi,{\mu}_2(\Psi, e))
	\Big)\,.\\
	&\hspace{0.2cm}\vdots\nonumber
	\end{align}
\end{subequations}
\endgroup
Going to the large Hilbert space and using cyclicity of the gauge products $\mu_k$, we obtain the following perturbative expression for the observable
\begin{align}
\mathcal{E}(\Psi) &= -\omega_\text{L}\big(e,\xi_0\Psi\big)-\frac{1}{2}\omega_\text{L}\big(e,\mu_2(\xi_0\Psi,\Psi)+\mu_2(\Psi,\xi_0\Psi)\big)+\nonumber\\
&\hspace{2cm}-\frac{1}{6}\omega_\text{L}\big(e,{\mu}_3(\xi_0\Psi, \Psi,\Psi)+{\mu}_3(\Psi,\xi_0\Psi,\Psi)+{\mu}_3( \Psi,\Psi,\xi_0\Psi)\big)+\nonumber\\
&\hspace{2cm}-\frac{1}{6}\omega_\text{L}\big(e,	{\mu}_2({\mu}_2(\xi_0\Psi,\Psi)+{\mu}_2(\Psi,\xi_0\Psi),\Psi)
+\nonumber\\
&\hspace{4cm}
+{\mu}_2(\Psi,{\mu}_2(\xi_0\Psi,\Psi)+{\mu}_2(\Psi, \xi_0\Psi))
+\nonumber\\[+1mm]
&\hspace{4cm}
+{\mu}_2({\mu}_2(\Psi,\Psi),\xi_0\Psi)
+{\mu}_2(\xi_0\Psi,{\mu}_2(\Psi,\Psi))\big)+\nonumber\\
&\hspace{2cm}+\mathcal{O}(\Psi^4)\,.\label{eq:Eexpanded}
\end{align}

\subsection{Relation with the Ellwood invariant of the WZW theory}

Let us now make use of the field redefinition \eqref{eq:fredef} of \cite{Erler:2015rra,Erler:2015uba} to investigate how the observable $\mathcal{E}(\Psi)$ of the Munich $A_\infty$ SFT, constructed in the previous subsection, relates to the Ellwood invariant in the Berkovits WZW-like theory. Rewriting first the observable $\mathcal{E}(\Psi)$ using the form \eqref{eq:obsMunich} in the large Hilbert space and substituting for $\mathbf{E}$ from \eqref{eq:GeG}, we obtain
\begin{align}
\mathcal{E}(\Psi) = -\int_0^1 dt\, \langle \omega_\text{L} |\pi_1 \bs{\xi}_t\frac{1}{1-\Psi(t)}\otimes \pi_1\mathbf{G}^{-1}\mathbf{e} \mathbf{G}\frac{1}{1-\Psi(t)}\,,\label{eq:IntStep}
\end{align}
where $\bs{\xi}_t$ denotes the coderivation associated to the 1-product $\xi_0 \p_t$.
Using then cyclicity of $\mathbf{G}$, recognizing the expression \eqref{eq:Adef2} for the WZW-like gauge potential component $\tilde{A}_t$ in terms of the $A_\infty$ SFT string field $\Psi(t)$, as well as using that 
\begin{align}
\pi_1\mathbf{e} \mathbf{G}\frac{1}{1-\Psi(t)}=\pi_1\mathbf{e}=e\,,
\end{align}
the expression \eqref{eq:IntStep} can be immediately recast in terms of the partially gauge-fixed WZW-like string field $\tilde\Phi(t)=\xi\tilde{\Psi}(t)$ as
\begin{align}
\mathcal{E}(\Psi) = -\int_0^1 dt\, \langle \omega_\text{L} | \Big[(\p_t e^{\xi_0\tilde{\Psi}(t)})e^{-\xi_0\tilde{\Psi}(t)}+\Delta A_t(t)\Big]\otimes {e}\,,
\end{align}
where the string field $\Delta A_t(t)$ satisfies the relation \eqref{eq:flatDA}. At this point, we note that the Ellwood state $e$ can be rewritten in a manifestly $\eta_0$-exact form as
\begin{align}
e=\eta_0 f\,,
\end{align}
where the state $f$ is again a local midpoint insertion on the identity string field.\footnote{
Recalling \eqref{eq:eNSNS} and using $\eta_0 Q\mathcal{U}^a=0$, in the NSNS sector we obtain $e = \eta_0 f_\text{NSNS}$ with
\begin{align}
f_\text{NSNS}=\tilde V_{ab}\left[ \mathcal{U}^a(i)\, Q \mathcal{U}^b(-i)-Q \mathcal{U}^a(i) \,  \mathcal{U}^b(-i)\right]I\,,
\end{align}
while in the RR sector we obtain $e = \eta_0 f_\text{RR}$ with (see \eqref{eq:eRR})
\begin{align}
f_\text{RR}=\tilde V_{\alpha\bar{\beta}}\left[
\xi cS^{\alpha}e^{-\phi/2}(i)\,\bar{c}\bar{S}^{\bar{\beta}}e^{-\bar{\phi}/2}(-i)
+cS^{\alpha}e^{-\phi/2}(i)\,\xi{c}{S}^{\bar{\beta}}e^{-{\phi}/2}(-i)
\right]I\,.
\end{align}
}
This means that we can rewrite
\begin{subequations}
\begin{align}
-\int_0^1 dt\, \langle \omega_\text{L} | \Delta A_t(t)\otimes {e}&=-\int_0^1 dt\, \langle \omega_\text{L} | \Delta A_t(t)\otimes D_\eta(t) f\\
&=-\int_0^1 dt\, \langle \omega_\text{L} | D_\eta(t)\Delta A_t(t)\otimes  f\\
&=0\,.
\end{align}
\end{subequations}
Here, in the first equality, we have used the fact that $\eta_0$ acting on $f$ can be replaced with the full action of the covariant dervative $D_\eta(t)$ (see \eqref{eq:covEta}), because $f$, being a local midpoint insertion on the identity string field commutes with $\tilde{A}_\eta(t)$. In the second equality, we have used the BPZ property of $D_\eta(t)$ (which follows from the BPZ property of $\eta_0$ and cyclicity of the star product $m_2$), while in the last equality, we have finally used the relation \eqref{eq:flatDA} satisfied by $\Delta A_t(t)$. This means that the term containing $\Delta A_t(t)$ actually does not contribute to $\mathcal{E}(\Psi)$, so that we eventually obtain
\begin{align}
\mathcal{E}(\Psi) = -\int_0^1 dt\, \langle \omega_\text{L} |(\p_t e^{\xi_0\tilde{\Psi}(t)})e^{-\xi_0\tilde{\Psi}(t)}\otimes {e}=\aver{e,\tilde\Phi}_L\,.
\end{align}
Hence, we note that under the field redefinition \eqref{eq:fredef} of \cite{Erler:2015rra,Erler:2015uba}, the proposed observable for the Munich open SFT naturally maps to the $t$-Ellwood invariant of the partially gauge-fixed Berkovits' open SFT (which we reviewed in Section \ref{sec:3}).

\section{Open-closed couplings in $A_\infty$ open superstring field theory}\label{sec:5}

The subject of this section will be to discuss tree-level effective actions obtained by classically integrating out massive degrees of freedom from the $A_\infty$ action for open superstring field theory, which is augmented by adding a term proportional the observable $\mathcal{E}(\Psi)$ introduced in Section \ref{sec:MunichObs}. The proportionality coupling will be denoted by $\mu$. We will see that the interactions of such a microscopic theory are given by the products $M_k^{(\mu)}$, which formally satisfy the relations of a weak $A_\infty$ algebra.
Such a theory is conjectured to be the open superstring field theory on a D-brane sitting in a closed string background deformed by turning on vev of the on-shell closed string field which enters the bosonic Ellwood state $e$. This will be put to a leading-order test later in the third paper  \cite{OC-III} of the series. 
There we will also focus on relating the open-closed effective couplings obtained from the microscopic WZW-like and $A_\infty$ actions for the first couple of orders in perturbation theory. In the spirit of the analysis presented in \cite{Maccaferri:2019ogq}, we find that the equality of these couplings at a fixed order in perturbation theory is guaranteed by imposing certain ``projector conditions'', which ensure flatness of the effective potential at all lower orders.

\subsection{Effective open-closed couplings to all orders}
\label{subsec:EffAinf}

Let us start by noting that as a consequence of the formal nilpotency \eqref{eq:nilp} of the coderivation $\mathbf{E}$ and also of the fact that we have $[\mathbf{E},\mathbf{M}]=0=[\mathbf{M},\mathbf{M}]$, one may deform the $A_\infty$ algebra of the superstring products $M_k$ by defining new (weak\footnote{By weak (or curved) $A_\infty$ products we mean an algebra of products $D_k$ encoded in an odd coderivation $\mathbf{D}$ such that $[\mathbf{D},\mathbf{D}]=0$ and $D_0\neq 0$.})
$A_\infty$ products $M_k^{(\mu)}\equiv \pi_1 \mathbf{M}^{(\mu)}\pi_k$, where we have introduced the $\mu$-deformed coderivation
\begin{align}
\mathbf{M}^{(\mu)}\equiv\mathbf{M}+\mu\mathbf{E}\label{eq:Mmu}
\end{align}
for a continuous parameter $\mu$. 
It is then easy to see that one can recast the action
\begin{align}
S^{(\mu)}(\Psi) = S(\Psi)+\mu\mathcal{E}(\Psi)\,,
\end{align}
obtained by deforming the $A_\infty$ open SFT action $S(\Psi)$ with the observable $\mathcal{E}(\Psi)$, as a (weak) $A_\infty$ action
\begin{align}
S^{(\mu)}(\Psi)=\sum_{k=0}^\infty \frac{1}{k+1}\omega_\text{S}(\Psi,M_{k}^{(\mu)}(\Psi^{\otimes k}))\,.
\end{align}
The products $M_k^{(\mu)}$ can be alternatively expressed as
\begin{align}
M_k^{(\mu)}=\left\{
\begin{array}{ll}
\mu E_0  & \text{for $k=0$}\\
M_k+ \mu E_k & \text{for $k>0$}
\end{array}
\right.\,.
\end{align}
Introducing for $0\leqslant t\leqslant 1$ an interpolation $\Psi(t)$ such that $\Psi(0)=0$, $\Psi(1)=\Psi$, we can therefore recast the deformed action $S^{(\mu)}(\Psi)$ in the coalgebra language as
\begin{align}
S^{(\mu)}(\Psi)=\int_0^1 dt\, \langle \omega_\text{S}|\pi_1\bs{\p}_t\frac{1}{1-\Psi(t)} \otimes \pi_1 \mathbf{M}^{(\mu)}\frac{1}{1-\Psi(t)}\,.
\end{align}
As usual, the fact that the products $M_k^{(\mu)}$ satisfy a (weak) $A_\infty$ algebra guarantees gauge symmetry for such an action.
As in the case of the cubic OSFT (see \cite{Erbin:2020eyc}), it is natural to conjecture that the action $S^{(\mu)}(\Psi)$ based on the new products $M_k^{(\mu)}$ describes open SFT with a marginal closed string background (determined by the on-shell closed-string state $e$) being turned on. Note that the new theory contains a tadpole term (given by the 0-string product $e$) which has to be removed by shifting the vacuum in order to restore the canonical $A_\infty$ form of the action (without a 0-string product). As it was originally discussed in \cite{Erbin:2020eyc} and \cite{OC-I} in the case of the bosonic string and in the previous sections in the case of the superstring in the large Hilbert space, the fact that this shift can be obstructed may be partially a manifestation of the background D-brane system being unable to adapt  to the bulk marginal deformation, as well as of the fact that the bulk deformation itself may not be exactly marginal.

We will now be interested in the algebraic aspects of integrating out the massive part $\mathcal{R}=(1-P_0)\Psi$ of the dynamical string field $\Psi$, where $P_0$ denotes the projector onto $\text{ker}\,L_0$. In practice, this is done by solving the equations of motion for $\mathcal{R}$ in terms of the remaining degrees of freedom $\psi=P_0\Psi$ (obtaining a solution $\mathcal{R}_\mu(\psi)$) and substituting the string field $\Psi_\mu(\psi)\equiv \psi+\mathcal{R}_\mu(\psi)$ back into the microscopic action $S^{(\mu)}(\Psi)$. Doing so, we end up with an effective action $\tilde{S}^{(\mu)}(\psi)=S^{(\mu)}(\Psi_{\mu}(\psi))$ for $\psi $. The solution $\mathcal{R}_{\mu}(\psi)$ for $\mathcal{R}$ takes the form of a tree-level Feynman diagram expansion in terms of the Siegel-gauge propagator $h = (b_0/L_0)\bar{P}_0$ satisfying the Hodge-Kodaira decomposition
\begin{align}
hQ+Qh=1-P_0\,,\label{eq:HK}
\end{align}
as well as the ``annihilation'' conditions $hP_0=P_0h=h^2=0$. This procedure is known to be automatically taken care of by the machinery of the homological perturbation lemma \cite{Erbin:2020eyc}: following the application of the homotopy transfer for the ($\mu$-deformed) interactions
$\delta\mathbf{M}^{(\mu)}\equiv\mathbf{M}^{(\mu)}-\mathbf{Q}$, we find that the solution for $\mathcal{R}_\mu(\psi)$ gives
\begin{align}
\Psi_\mu(\psi)\equiv \pi_1\frac{1}{\mathbf{1}_{T\mathcal{H}}+\mathbf{h}\delta\mathbf{M}^{(\mu)}}\mathbf{I}_0\frac{1}{1_{T\mathcal{H}}-\psi}\,,\label{eq:Psimupsi}
\end{align}
where $1_{T\mathcal{H}}$ is the identity element of $T\mathcal{H}$. We also introduce the cohomomorphisms $\bs{\Pi}_0$, $\mathbf{I}_0$ corresponding to the canonical projection $\Pi_0:\mathcal{H}\longrightarrow P_0\mathcal{H}$ and the canonical inclusion $I_0:P_0\mathcal{H}\longrightarrow \mathcal{H}$ (i.e.\ we have $\bs{\Pi}_0\pi_k = \Pi_0^{\otimes k}$, $\mathbf{I}_0\pi_k = I_0^{\otimes k}$), $\mathbf{1}_{T\mathcal{H}}$ is the identity cohomomorphism on $T\mathcal{H}$ and $\mathbf{h}$ is the lift of the propagator $h$ to a map on $T\mathcal{H}$ defined such that the lifted Hodge-Kodaira decomposition
\begin{align}
\mathbf{h}\mathbf{Q}+\mathbf{Q}\mathbf{h}=\mathbf{1}_{T\mathcal{H}}-\mathbf{I}_0\bs{\Pi}_0
\end{align}
holds together with the annihilation conditions $\mathbf{\Pi}_0\mathbf{h}=\mathbf{h}\mathbf{I}_0=\mathbf{h}^2=0$. Substituting the string field $\Psi_\mu(\psi)$ into the microscopic action $S^{(\mu)}(\Psi)$, the classical effective action $\tilde{S}^{(\mu)}(\psi)=S^{(\mu)}(\Psi_\mu(\psi))$ can be expressed as a weak $A_\infty$ action
\begin{align}
\tilde{S}^{(\mu)}(\psi)=S^{(\mu)}(\Psi_\mu(0))+\int_0^1 dt\, \langle \tilde{\omega}_\text{S}|\pi_1\bs{\p}_t\frac{1}{1-\psi(t)} \otimes \pi_1 \tilde{\mathbf{M}}^{(\mu)}\frac{1}{1-\psi(t)}\,,\label{eq:EffAct}
\end{align}
based on the products $\tilde{M}_k^{(\mu)}\equiv \pi_1 \tilde{\mathbf{M}}^{(\mu)}\pi_k$, where
\begin{align}
\tilde{\mathbf{M}}^{(\mu)}=\bs{\Pi}_0\mathbf{M}^{(\mu)}\frac{1}{\mathbf{1}_{T\mathcal{H}}+\mathbf{h}\delta\mathbf{M}^{(\mu)}}\mathbf{I}_0\,.
\end{align}
Here the symplectic form $\tilde{\omega}_\text{S}$ is defined by restricting $\omega_\text{S}$ to $P\mathcal{H}$ and the interpolation $\psi(t)$ runs from $\psi(0)=0$ to $\psi(1)=\psi$. Also note that the constant term $S^{(\mu)}(\Psi_\mu(0))$ has no effect on the dynamics.
For more details about the coalgebra notation, we refer the reader to ref.\ \cite{Erbin:2020eyc}.

Applying the vertical decomposition procedure described in Section 2.5 of \cite{Erbin:2020eyc}, the coderivation $\tilde{\mathbf{M}}^{(\mu)}$ can be recast as an explicit perturbation series in $\mu$, that is
\begin{align}
\tilde{\mathbf{M}}^{(\mu)} = \sum_{\alpha=0}^\infty \mu^\alpha\tilde{\mathbf{N}}_\alpha\,,
\end{align}
where we have introduced the coderivations 
\begin{align}
\tilde{\mathbf{N}}_\alpha =\left\{
\begin{array}{ll}
\mathbf{\Pi}_0\mathbf{M}\frac{1}{\mathbf{1}_{T\mathcal{H}}+\mathbf{h}\delta\mathbf{M}}\mathbf{I}_0 & \text{for $\alpha=0$}\\
\mathbf{\Pi}_0\frac{1}{\mathbf{1}_{T\mathcal{H}}+\delta\mathbf{M}\mathbf{h}}\mathbf{E}\bigg[\!-\!\frac{1}{\mathbf{1}_{T\mathcal{H}}+\mathbf{h}\delta\mathbf{M}}\mathbf{h}\mathbf{E}\bigg]^{\alpha-1}\frac{1}{\mathbf{1}_{T\mathcal{H}}+\mathbf{h}\delta\mathbf{M}}\mathbf{I}_0 & \text{for $\alpha>0$}
\end{array}
\right.\,,\label{eq:Ncoder}
\end{align}
with $\delta\mathbf{M}\equiv \delta \mathbf{M}^{(\mu=0)} \equiv \mathbf{M}-\mathbf{Q}$ being the interacting part of the undeformed coderivation $\mathbf{M}$. Denoting the homotopy transfer of $\mathbf{M}$, $\mathbf{E}$ for the undeformed interactions $\delta \mathbf{M}$ by
\begin{subequations}
	\begin{align}
	\tilde{\mathbf{M}}&=\bs{\Pi}_0\frac{1}{\mathbf{1}_{T\mathcal{H}}+\delta\mathbf{M}\mathbf{h}}\mathbf{M}\frac{1}{\mathbf{1}_{T\mathcal{H}}+\mathbf{h}\delta\mathbf{M}}\mathbf{I}_0\equiv\tilde{\mathbf{N}}_0\,,\\
	\tilde{\mathbf{E}}&=\bs{\Pi}_0\hspace{0.5mm}\frac{1}{\mathbf{1}_{T\mathcal{H}}+\delta\mathbf{M}\mathbf{h}}\mathbf{E}\hspace{0.5mm}\frac{1}{\mathbf{1}_{T\mathcal{H}}+\mathbf{h}\delta\mathbf{M}}\hspace{0.5mm}\mathbf{I}_0\equiv\tilde{\mathbf{N}}_1\,,\label{eq:Etilde}
	\end{align}
\end{subequations}
we can write
\begin{align}
\tilde{\mathbf{M}}^{(\mu)} = \tilde{\mathbf{M}}+\mu\tilde{\mathbf{E}}+\mathcal{O}(\mu^2)\,.\label{eq:Mmutilde}
\end{align}
Compared to \eqref{eq:Mmu}, it is therefore necessary to add the higher order corrections $\mathcal{O}(\mu^2)$ in \eqref{eq:Mmutilde}, which then conspire to restore the nilpotency of the coderivation $\tilde{\mathbf{M}}^{(\mu)}$ (and therefore the gauge invariance of the effective action $\tilde{S}^{(\mu)}(\psi)$). Noting that the relation $[\mathbf{E},\mathbf{M}]=0$ implies an analogous relation
\begin{align}
[\tilde{\mathbf{E}},\tilde{\mathbf{M}}]=0
\end{align}
for the homotopy transfer of the coderivation $\mathbf{E}$ and $\mathbf{M}$ (with interactions $\delta\mathbf{M}$), we can conclude (see Section 2.3 of \cite{Erbin:2020eyc}) that the homotopy transfer
\begin{align}
\tilde{\mathcal{E}}(\psi)
\equiv \int_0^1 dt\,\langle\tilde{\omega}_\text{S} |\, \pi_1\bs{\p}_t \frac{1}{1-\psi(t)}\otimes \pi_1 \tilde{\mathbf{E}} \frac{1}{1-\psi(t)}\equiv \mathcal{E}(\Psi(\psi))
\end{align}
of $\mathcal{E}(\Psi)$ is an observable of the effective theory with (undeformed) products $\tilde{M}_k = \pi_1\tilde{\mathbf{M}}\pi_k$. The relation \eqref{eq:Mmutilde} therefore says that simply deforming the effective action $\tilde{S}(\psi)$ (based on the $A_\infty$ products $\tilde{M}_k$) with the observable $\tilde{\mathcal{E}}(\psi)$ does not yield a consistent SFT action and one has to add higher $\mu$ corrections so as to end up with a weak-$A_\infty$ action $\tilde{S}^{(\mu)}(\psi)$. Ignoring the constant term, this can be expanded order by order in $\mu$ as
\begin{align}
\tilde{S}^{(\mu)}(\psi) = \sum_{k=0}^\infty\sum_{\alpha=0}^\infty \mu^\alpha\tilde{S}_{k,\alpha}(\psi)\,,
\end{align}
where we have denoted by $\tilde{S}_{k,\alpha}(\psi)$ the vertices which couple $k+1$ open-string insertions with $\alpha$ (on-shell) closed-string insertions. These vertices can be expressed explicitly in terms of the products $\tilde{N}_{k,\alpha} \equiv \pi_1\tilde{\mathbf{N}}_\alpha\pi_k$
as
\begin{align}
\tilde{S}_{k,\alpha}(\psi)=\frac{1}{k+1}\omega_\text{S}(\psi,\tilde{N}_{k,\alpha}(\psi^{\otimes k}))\,.\label{eq:Ska}
\end{align}
Nilpotency of the coderivation $\tilde{\mathbf{M}}^{(\mu)}$ then implies that the products $\tilde{N}_{k,\alpha}$ satisfy the algebra
\begin{align}
\sum_{l=0}^{k}\sum_{\beta=0}^{\alpha}[\tilde{N}_{l,\beta},\tilde{N}_{k-l,\alpha-\beta}]=0
\end{align}
for $k,\alpha\geqslant 0$ (where we set $\tilde{N}_{0,0}=0$). Evaluating explicitly the products $\tilde{N}_{k,\alpha}$ for a couple of lowest orders using \eqref{eq:Ncoder}, we obtain
\begin{subequations}
	\begin{align}
	\tilde{N}_{0,1}&=P_0e\,,\\
	\tilde{N}_{0,2}&=-P_0E_1(he)+P_0M_2(he,he)\,,\\
	\tilde{N}_{1,0}(\psi)&=P_0Q\psi\,,\\
	\tilde{N}_{1,1}(\psi)&=P_0E_1(\psi)-P_0M_2(he,\psi)-P_0M_2(\psi,he)\,,\\
	\tilde{N}_{2,0}(\psi,\psi)&=P_0M_2(\psi,\psi)\,,\\
	\tilde{N}_{3,0}(\psi,\psi)&=P_0M_3(\psi,\psi,\psi)-P_0M_2(\psi,hM_2(\psi,\psi))-P_0M_2(hM_2(\psi,\psi),\psi)\,.\\
	&\hspace{0.2cm}\vdots\nonumber
	\end{align}
\end{subequations}
Substituting these into \eqref{eq:Ska}, we obtain the corresponding effective action vertices, which we expect to be related by a field redefinition to the WZW-like effective interactions \eqref{eq:SeffWZW}. While we will not attempt to characterize this field redefinition, we will verify in the final paper \cite{OC-III} of the series (for a couple of lowest orders) that the WZW-like effective potential vanishes if and only if the $A_\infty$ effective potential vanishes, namely that the two theories yield the same constraints on the moduli for any given background.

It is important to note that any classical solution of the effective theory yields a classical solution of the microscopic theory. Indeed, as it was reviewed in detail in \cite{Erbin:2020eyc}, given a classical solution $\psi^\ast$ of the equations of motion derived by varying the effective action $\tilde{S}^{(\mu)}(\psi)$, the string field $\Psi^\ast \equiv \Psi_\mu(\psi^\ast)$ solves the microscopic equation of motion. 
An application of this observation arises when discussing the vacuum shift which one needs to perform in order to remove the tadpoles from both the microscopic and the effective action. Indeed, the couplings $\tilde{S}_{0,\alpha}(\psi)$ signal the presence of a tadpole in the effective action, which originates	from the tadpole of the microscopic theory given by the 0-string product $E_0$. As it was extensively discussed in \cite{Erbin:2020eyc} for the bosonic OSFT, the tadpoles of both the microscopic and the effective theory can be removed by expanding the action around the string fields $\Psi_\mu\in \mathcal{H}$ and $\psi_\mu\in P_0\mathcal{H}$, respectively (with the property $\Psi_{\mu=0}=\psi_{\mu=0}=0$), where the equations of motion
\begin{align}
\pi_1 \tilde{\mathbf{M}}^{(\mu)}\frac{1}{1_{{T}\mathcal{H}}-\psi_\mu}=0
\end{align}
for the effective vacuum shift $\psi_\mu$ can be interpreted as obstructions to solving the equations of motion for the microscopic vacuum shift $\Psi_\mu$. Put in other words, the tadpole of both the microscopic and the effective action can be removed if and only if the equations of motion for $\psi_\mu$ can be solved. The corresponding microscopic vacuum shift is then simply given  as
\begin{align}
\Psi_\mu= \pi_1\frac{1}{\mathbf{1}_{T\mathcal{H}}+\mathbf{h}\delta\mathbf{M}^{(\mu)}}\mathbf{I}_0\frac{1}{1_{T\mathcal{H}}-\psi_\mu}\equiv \Psi_\mu(\psi_\mu)\label{eq:MicVacShift}
\end{align}
in terms of the effective vacuum shift $\psi_\mu$. Assuming that the effective vacuum shift $\psi_\mu$ can be consistently set to zero, the microscopic vacuum shift can then be expanded as
\begin{align}
\Psi_\mu = -\mu he+\mu^2 h \big[E_1(he)-M_2(he,he)\big]+\mathcal{O}(\mu^3)\,.
\end{align}

\section{Discussion and outlook}\label{sec:7}
In this paper we have extended the computation of the effective open-closed couplings originally derived for the bosonic string in \cite{Erbin:2020eyc, Koyama:2020qfb} and later discussed in \cite{OC-I}, to open superstring field theory in the NS sector in both the large and small Hilbert space. Leaving for the third paper \cite{OC-III} the discussion of the relation between the two effective field theories, and the concrete calculation of some non trivial open-closed coupling, here we offer some thoughts on possible future research related to the results we have presented in this paper
\begin{itemize}
\item Given the fact that the WZW theory is amenable to exact analytic techniques which are natural extensions of the bosonic ones (see for example \cite{Noumi:2011kn, Erler:2013wda}) it would be natural to search for analytic vacuum shift solutions. The situation is perhaps a little less promising than in the bosonic case as the status of exact superstring solutions is still in its infancy and so many natural classical solutions (for example lower dimensional BPS branes obtained via tachyon condensation from unstable higher dimensional systems) are still to be found. In this regard the search for analytic vacuum shift solutions could be an interesting new direction towards analytic methods. Ideally, we would like to describe all possible D-brane systems which are compatible with a given closed string background as classical solutions generalizing  \cite{Erler:2019nmz, Erler:2019fye, Erler:2014eqa} and then we would like to be able to determine their deformations induced by the Ellwood invariant. 

\item In the bosonic OSFT the Ellwood invariant  provided a rather direct way to the computation of the full boundary state defined by a given solution \cite{Kudrna:2012re}. This construction has not been extended to the superstring and it would be interesting to do so.

\item We have observed that the WZW theory in the large Hilbert space has a peculiar perturbation theory which is manifestly symmetric in $(Q,\eta)$ and in the associated propagators $(h,\tilde h)$. It would be interesting to study the systematics of this perturbation theory to understand the emergence of the effective theory in the large Hilbert space (of which we have computed only the first few terms). In the $A_\infty$ small Hilbert space theories this is provided by the homotopy transfer, but in WZW case it seems we lack of an analogous  convenient packaging of the perturbation theory.

\item It would be clearly instructive to complete our analysis by adding the Ramond sector.

\end{itemize}
We hope that progress in the above directions will be possible in the upcoming future.

\section*{Acknowledgments}
We thank Ted Erler and Martin Schnabl for discussions and Ashoke Sen for correspondence.
We thank the organizers of ``Fundamental Aspects of String Theory'', Sao Paolo 1-12 June 2020,  and in particular Nathan Berkovits for giving us the opportunity to present some of our results prior to publication.
CM thanks CEICO and the Czech Academy of Science for hospitality during part of this work. JV also thanks INFN Turin for their hospitality during the initial stages of this work.  
The work of JV was supported by the NCCR SwissMAP that is funded by the Swiss National Science Foundation.
The work of CM is partially supported by the MIUR PRIN
Contract 2015MP2CX4 ``Non-perturbative Aspects Of Gauge Theories And Strings''.

\appendix

\section{Review of the  $A_\infty$ construction in the small Hilbert space}
\label{app:MunichReview}

Here we will give a brief review of the way the products of the $A_\infty$ open superstring field theory are built. See the original papers \cite{Erler:2013xta,Erler:2014eba} for more details. We will also remind the reader of the field redefinition \cite{Erler:2015rra,Erler:2015uba} mapping this theory to the partially gauge-fixed WZW-like Berkovits open SFT. For a review of the tensor coalgebra notation, which we will heavily rely upon, see for instance \cite{Erbin:2020eyc}.

\subsection{Product notation}

The action of the ``Munich'' $A_\infty$ open superstring field theory can be written as
\begin{align}
{S}(\Psi) = \sum_{k=1}^\infty\frac{1}{k+1}\omega_\text{S}(\Psi,M_k(\Psi^{\otimes k}))\,,\label{eq:actMunich}
\end{align}
where $\Psi$ is the dynamical string field in the small Hilbert space $\mathcal{H}_\text{S}$ at ghost number $1$ and picture number $-1$. In the following, we will use grading by degree $d(A)=|A|+1$, where $|A|$ denotes the ghost number of $A$. The dynamical string field is therefore degree-even. The degree-odd small Hilbert space \emph{superstring products} $M_k:(\mathcal{H}_\text{S})^{\otimes k}\longrightarrow \mathcal{H}_\text{S}$ carrying picture number $k-1$ and ghost number $2-k$ satisfy the $A_\infty$ relations
\begingroup\allowdisplaybreaks
\begin{subequations}
	\label{eq:ainf}
	\begin{align}
	0&=M_1(M_1(A_1))\,,\label{eq:nil}\\
	0&=M_1(M_2(A_1,A_2))+M_2(M_1(A_1),A_2)+\nonumber\\
	&\hspace{4cm}+(-1)^{d(A_1)}M_2(A_1,M_1(A_2))\,,\label{eq:der}\\
	0&=M_1(M_3(A_1,A_2,A_3))+M_2(M_2(A_1,A_2),A_3)+\nonumber\\
	&\hspace{4cm}+(-1)^{d(A_1)}M_2(A_1,M_2(A_2,A_3))+\nonumber\\
	&\hspace{1cm}+M_3(M_1(A_1),A_2,A_3)+(-1)^{d(A_1)}M_3(A_1,M_1(A_2),A_3)+\nonumber\\
	&\hspace{4cm}+(-1)^{d(A_1)+d(A_2)}M_3(A_1,A_2,M_1(A_3))\,,\label{eq:ass}\\
	&\hspace{0.2cm}\vdots\nonumber
	\end{align}
\end{subequations}
\endgroup
which provide the non-linear gauge invariance of $S(\Psi)$.
These can be constructed in terms of the degree-odd \emph{bosonic products} $m_1\equiv Q\equiv M_1$ and $m_2$ (the Witten's star product) by going through intermediate steps in the large Hilbert space, the so-called \emph{gauge products} $\mu_k$, which are degree-even. Finally, $\langle \omega_\text{S}|:(\mathcal{H}_\text{S})^{\otimes 2}\longrightarrow \mathbb{C}$ denotes the symplectic (graded-antisymmetric) form on the small Hilbert space which is defined in terms of the BPZ product $\langle\cdot,\cdot\rangle$ and with respect to the products $M_k$ are cyclic. That is, we have
\begin{align}
\omega_\text{S}(\Psi_1,\Psi_2)&=-(-1)^{d(\Psi_1)}\langle\Psi_1,\Psi_2\rangle_\mathrm{S}\,,\label{eq:symplBPZ}
\end{align}
as well as
\begin{subequations}
	\begin{align}
	\omega_\text{S}(A_1,A_2) &=-(-1)^{d(A_1)d(A_2)}\omega_\text{S}(A_2,A_1)\,,\\
	\omega_\text{S}(A_1,M_k(A_2,\ldots,A_{k+1}))&=-(-1)^{d(A_1)}\omega_\text{S}(M_k(A_1,\ldots,A_k),A_{k+1})\,.
	\end{align}	
\end{subequations}
Varying the action \eqref{eq:actMunich}, we obtain the equation of motion
\begin{align}
0 = \sum_{k=1}^\infty M_k(\Psi^{\otimes k})\,.
\end{align}
Furthermore, the action is invariant under the gauge transformation
\begin{align}
\delta_\Lambda \Psi = \sum_{k=1}^\infty\sum_{l=0}^{k-1}M_k(\Psi^{\otimes l},\Lambda,\Psi^{\otimes (k-1-l)})\,.\label{eq:gaugeT}
\end{align}
We now turn to the definition of the products $M_k$. Setting $M_1\equiv Q$ and $\mu_1\equiv 0$, one can define the superstring 2-product $M_2$ either directly by smearing the PCO zero-mode $X_0$ cyclically over the insertions of $m_2$ as
\begin{align}
M_2(A_1,A_2)\equiv \frac{1}{3}\Big[X_0 m_2(A_1,A_2)+m_2(X_0 A_1,A_2)+m_2(A_1,X_0 A_2)\Big]\,,
\end{align}
(so that it is manifest that $M_2$ is in the small Hilbert space), or equivalently, in terms of the large Hilbert space gauge 2-product by first graded-smearing the superghost zero-mode $\xi_0$ cyclically over the $m_2$ insertions 
\begin{align}
\mu_2(A_1,A_2) \equiv \frac{1}{3}\Big[\xi_0 m_2(A_1,A_2)-m_2(\xi_0 A_1,A_2)-(-1)^{d(A_1)}m_2(A_1,\xi_0A_2)\Big]\label{eq:mudef}
\end{align}
and then putting
\begin{align}
M_2 = [Q,\mu_2]\,.
\end{align}
Such a definition yields $M_2$ which is non-associative and one therefore needs to define a 3-string product $M_3$ in order to restore gauge invariance of the action. For the purposes of constructing the superstring products $M_k$ for $k\geq 3$, it will prove beneficial to uplift the products to coderivations acting on a tensor coalgebra, which we shall now explain.

\subsection{Tensor coalgebra construction}
\label{app:coalg}

Introducing the degree-odd cyclic coderivations $\mathbf{M}_k$ corresponding to the products $M_k$ which act on the tensor space 
\begin{align}
T\mathcal{H}_\text{S} = (\mathcal{H}_\text{S})^{\otimes 0}\oplus(\mathcal{H}_\text{S})^{\otimes 1}\oplus(\mathcal{H}_\text{S})^{\otimes 2}\oplus\ldots
\end{align}
equipped with the usual deconcatenation coproduct (thus forming a tensor coalgebra; see \cite{Erbin:2020eyc} for more details),
we define the total (cyclic) coderivation $\mathbf{M}$ as $\mathbf{M}=\sum_{k=1}^\infty \mathbf{M}_k$. We can therefore recover the products $M_k$ from $\mathbf{M}$ as $M_k = \pi_1\mathbf{M}\pi_k$.
The $A_\infty$ relations \eqref{eq:ainf} satisfied by $M_k$ can then be neatly expressed as
\begin{align}
[\mathbf{M},\mathbf{M}]=0\,.\label{eq:ainfCoder}
\end{align}
Introducing also an arbitrary interpolation $\Psi(t)$ for $0\leq t \leq 1$ such that $\Psi(0)=0$ and $\Psi(1)=\Psi$ and the coderivation $\bs{\p}_t$ corresponding to the operator $\p_t$ viewed as a 1-string product, the action \eqref{eq:actMunich} can be equivalently rewritten as
\begin{align}
S(\Psi)=\int_0^1 dt\,\langle\omega_\text{S} |\, \pi_1\bs{\p}_t \frac{1}{1-\Psi(t)}\otimes \pi_1 \mathbf{M} \frac{1}{1-\Psi(t)}\,.\label{eq:AinfAct}
\end{align}
The construction of higher superstring products $M_k$ can then be summarized in terms of the recurrent scheme\footnote{For any operator $O:\mathcal{H}\longrightarrow \mathcal{H}$ and a $k$-product $p_k:\mathcal{H}^{\otimes k}\longrightarrow \mathcal{H}$, the notation $O\circ p_k$ defines a $k$-product given by the graded cyclic smearing operation which already appeared in the cases $O=X_0,\xi_0$ and $p_2 = m_2$ in the previous subsection.}
\begin{subequations}
	\label{eq:MunichRecA2}
	\begin{align}
	\mathbf{M}_{n+1}^{(n-1)}&=\frac{1}{n-1}\bigg(
	[\mathbf{M}_2^{(0)},\boldsymbol{\mu}_{n}]
	+[\mathbf{M}_3^{(1)},\boldsymbol{\mu}_{n-1}]
	+\ldots
	+[\mathbf{M}_{n}^{(n-2)},\boldsymbol{\mu}_{2}]
	\bigg)\,,\label{eq:defM1}\\
	\bs{\mu}_{n+1}&=\xi_0\circ \mathbf{M}_{n+1}^{(n-1)}\,,\label{eq:defmu}\\[0.8mm]
	\mathbf{M}_{n+1}&=\frac{1}{n}\bigg([\mathbf{M}_1,\boldsymbol{\mu}_{n+1}]+[\mathbf{M}_2,\boldsymbol{\mu}_{n}]+\ldots+[\mathbf{M}_{n},\boldsymbol{\mu}_2]\bigg)\,,\label{eq:defM}
	\end{align}
\end{subequations} 
where initially we set ${M}_2^{(0)}\equiv m_2$ and $M_1\equiv Q$. First, it is manifest from the recursion relations \eqref{eq:MunichRecA2} that the coderivations $\mathbf{M}_k$, $\mathbf{M}_k^{(k-2)}$ are cyclic with respect to $\omega_\text{S}$ (and $\bs{\mu}_k$ is cyclic with respect to $\omega_\text{L}$) since $\mathbf{Q}$, $\mathbf{m}_2$ are cyclic with respect to $\omega_\text{S}$: this is true because 1. the commutator of two cyclic coderivations is again cyclic, 2. the cyclic smearing operation $\xi_0\circ$ manifestly preserves cyclicity and 3. given a small Hilbert space coderivation which is cyclic with respect to $\omega_\text{L}$, it is also automatically cyclic with respect to $\omega_\text{S}$. Defining $\mathbf{M}^{[1]}=\sum_{k=1}^\infty \mathbf{M}_k^{(k-2)}$, it can then be shown using the technique of generating functions\footnote{We can see a version of this argument in section \ref{sec:MunichObs} when we prove the recursion formula \eqref{eq:conjE} for the new observable $\mathbf{E}$.} that the recursion \eqref{eq:MunichRecA2} implies the relations \cite{Erler:2013xta,Erler:2014eba}
\begin{subequations}
	\begin{align}
	\mathbf{M} &= \mathbf{G}^{-1}\mathbf{Q}\mathbf{G}\,,\label{eq:GQG}\\
	\mathbf{M}^{[1]} &= \mathbf{G}^{-1}\mathbf{m}_2\mathbf{G}\,,
	\end{align}
\end{subequations}
where the cyclic cohomomorphism $\mathbf{G}$ can be written in terms of $\bs{\mu}(t) \equiv \sum_{k=0}^\infty t^k \bs{\mu}^{(k+1)}_{k+2}$ as a path-ordered exponential
\begin{align}
\mathbf{G} = \mathcal{P}_\tau\exp\bigg(\int_0^1 d\tau\,\bs{\mu}(\tau)\bigg)\label{eq:cohomG}
\end{align}
with the ordering prescription
\begin{align}
\mathcal{P}_\tau\big[\bs{\mu}(\tau_1)\bs{\mu}(\tau_2)\big]=\left\{\begin{array}{l}
\bs{\mu}(\tau_1)\bs{\mu}(\tau_2)\quad\text{if $\tau_1<\tau_2$}\\
\bs{\mu}(\tau_2)\bs{\mu}(\tau_1)\quad\text{if $\tau_1>\tau_2$}
\end{array}\right.\,.
\end{align}
We also have the relation \cite{Erler:2015rra}
\begin{align}
\mathbf{G}^{-1}\bs{\eta}\mathbf{G}=\bs{\eta}-\mathbf{m}_2\,.\label{eq:GetaG}
\end{align}
Using \eqref{eq:GetaG}, it is then straightforward to show that $[\bs{\eta},\mathbf{M}]=0=[\bs{\eta},\mathbf{M}^{[1]}]$, namely that the products $M_k$, $M_k^{(k-2)}$ are in the small Hilbert space. The form of \eqref{eq:GQG} also makes it manifest that $\mathbf{M}$ satisfies \eqref{eq:ainfCoder}, that is that the products $M_k$ satisfy $A_\infty$ relations.

\subsection{Relation to the WZW-like open SFT}

Let us now review the field redefinition mapping between the $A_\infty$- and the partially gauge-fixed WZW-like open superstring field theories which was originally established in \cite{Erler:2015rra,Erler:2015uba,Erler:2015uoa}. Substituting the expression \eqref{eq:GQG} into the $A_\infty$ SFT action \eqref{eq:AinfAct} and using cyclicity of the cohomomorphism $\mathbf{G}$ as well as the relation \eqref{eq:symplBPZ} between the symplectic form and the BPZ product, it is straightforward to rewrite the action in the form
\begin{align}
S(\Psi)&=-\int_0^1 dt\,\big\langle\tilde{A}_t(\Psi(t)),Q\tilde{A}_\eta(\Psi(t))\big\rangle\,,\label{eq:AinfBOZt}
\end{align}
where we have introduced string fields
\begin{subequations}
	\label{eq:Adef}
	\begin{align}
	\tilde{A}_\eta&=\pi_1 \mathbf{G}\frac{1}{1-\Psi(t)}\,,\\
	\tilde{A}_t&=\pi_1\mathbf{G}\bs{\xi}_t\frac{1}{1-\Psi(t)}\,.\label{eq:Adef2}
	\end{align}
\end{subequations}
One can then verify that these satisfy the relations
\begin{subequations}
	\begin{align}
	0&=\eta \tilde{A}_\eta - m_2(\tilde{A}_\eta ,\tilde{A}_\eta)\,,\label{eq:etaOSFT}\\
	0&=\eta \tilde{A}_t - \p_t \tilde{A}_\eta -m_2(\tilde{A}_t, \tilde{A}_\eta) - m_2(\tilde{A}_\eta ,\tilde{A}_t)\,,
	\end{align}
\end{subequations}
so that $\tilde{A}_t$ and $\tilde{A}_\eta$ can be interpreted as components of a flat connection. Note that since $\tilde{A}_\eta$ needs to solve the OSFT-like equation \eqref{eq:etaOSFT}, where the ``kinetic'' operator $\eta$ has trivial cohomology in the large Hilbert space (since $\xi_0$ acts as the corresponding contracting homotopy), we must be able to write it in a pure gauge form
\begin{align}
\tilde{A}_\eta = (\eta e^{\xi_0\tilde{\Psi}(t)})e^{-\xi_0\tilde{\Psi}(t)}
\end{align}
for some $\tilde{\Psi}(t)\in\mathcal{H}_\text{S}$. Since the form \eqref{eq:AinfBOZt} of the $A_\infty$ SFT action closely resembles the so-called dual form of the WZW-like action, it is therefore natural to introduce the field redefinition relating the string field $\Psi$ of the $A_\infty$ SFT to the string field $\tilde{\Psi}$ of the partially-gauged WZW-like open SFT
as follows
\begin{align}
\pi_1 \mathbf{G} \frac{1}{1-\Psi(t)} &= (\eta e^{\xi_0 \tilde{\Psi}(t)})e^{-\xi_0 \tilde{\Psi}(t)}\,.\label{eq:fredef}
\end{align}
It was then shown by \cite{Erler:2015rra} that given the relation \eqref{eq:fredef} between $\Psi(t)$ and $\tilde{\Psi}(t)$, the expression \eqref{eq:Adef} for $\tilde{A}_t$ in terms of $\Psi$ is then consistent with equating
\begin{align}
\tilde{A}_t =  (\p_t e^{\xi_0 \tilde{\Psi}(t)})e^{-\xi_0 \tilde{\Psi}(t)}+\Delta A_t(t)\,,
\end{align}
where the quantity $\Delta A_t(t)$ satisfies the relation
\begin{align}
0=D_\eta(t) \Delta A_t(t)\,.\label{eq:flatDA}
\end{align}
As usual, the covariant derivative $D_\eta(t)$ is defined as
\begin{align}
D_\eta(t)\Phi = \eta\Phi-A_\eta(t)\Phi+(-1)^{|\Phi|}\Phi A_\eta(t)\,.\label{eq:covEta}
\end{align}

\section{Detailed calculations}
\label{app:proofs}

Here we give some detailed calculations and proofs of various results used in the main body of this paper.
\subsection{Properties of $\mathbf{E}_2$}
\label{app:E2}

Let us give a detailed investigation of the properties of the coderivation $\mathbf{E}_2$, which we define recursively in terms of $\mathbf{E}_0$ and $\mathbf{E}_1$ as
\begin{align}
	\mathbf{E}_2 \equiv \frac{1}{2}\Big([\mathbf{E}_1,\boldsymbol{\mu}_2]+[\mathbf{E}_0,\boldsymbol{\mu}_3]\Big)\,.
\end{align}
Recall that we have already seen in Section \ref{sec:MunichObs} that $\mathbf{E}_0$ and $\mathbf{E}_1$ satisfy 
\begin{subequations}
	\label{eq:defE}
	\begin{align}
		\mathbf{E}_0 &\equiv \mathbf{e}\,,\\
		\mathbf{E}_1 &\equiv [\mathbf{E}_0,\bs{\mu}_2]\,,
	\end{align}
\end{subequations}
they are both in the small Hilbert space, that is
\begin{subequations}
	\label{eq:etaE}
\begin{align}
[\bs{\eta},\mathbf{E}_0]&=0\,,\\
[\bs{\eta},\mathbf{E}_1]&=0\,,
\end{align}
\end{subequations}
and they also satisfy the relation \eqref{eq:[E,M]k} for $k=1$ and $k=2$, namely
\begin{subequations}
	\begin{align}
	[\mathbf{E}_0,\mathbf{M}_1]&=0\,,\\
	[\mathbf{E}_0,\mathbf{M}_2]+[\mathbf{E}_1,\mathbf{M}_1]&=0\,.
	\end{align}
\end{subequations}
We will start by showing that $[\bs{\eta},\mathbf{E}_2]=0$. Indeed, using \eqref{eq:etaE}, we can first write
	\begin{align}
		[\bs{\eta},\mathbf{E}_2] =-\frac{1}{2}\Big([\mathbf{E}_1,[\bs{\eta},\boldsymbol{\mu}_2]]+[\mathbf{E}_0,[\bs{\eta},\boldsymbol{\mu}_3]]\Big)\,.
	\end{align}
Using the results
\begin{subequations}
\begin{align}
[\bs{\eta},\boldsymbol{\mu}_2]&=\mathbf{m}_2\,,\\
[\bs{\eta},\boldsymbol{\mu}_3]&=[\mathbf{m}_2,\bs{\mu}_2]\,,
\end{align}
\end{subequations}
and the definition \eqref{eq:defE} of $\mathbf{E}_1$, we can therefore write
\begin{subequations}
	\begin{align}
	[\bs{\eta},\mathbf{E}_2] &=\frac{1}{2}\Big([[\bs{\mu}_2,\mathbf{E}_0],\mathbf{m}_2]+[\mathbf{E}_0,[\boldsymbol{\mu}_2,\mathbf{m}_2]]\Big)\\
	&=\frac{1}{2}[\bs{\mu}_2,[\mathbf{E}_0,\mathbf{m}_2]]
\end{align}
\end{subequations}
which is zero because the bosonic Ellwood coderivation $\mathbf{e}$ of course satisfies $[\mathbf{e},\mathbf{m}_2]=0$. Let us continue by verifying the relation \eqref{eq:[E,M]k} for $k=3$. Using the relations
\begin{subequations}
	\begin{align}
		[\mathbf{M}_1,\bs{\mu}_2]&=\mathbf{M}_2\,,\\
		[\mathbf{M}_1,\bs{\mu}_3]&=2\mathbf{M}_3-[\mathbf{M}_2,\bs{\mu}_2]\,,
	\end{align}
\end{subequations}
we can first express
\begin{subequations}
\begin{align}
[\mathbf{M}_1,\mathbf{E}_2]&= \frac{1}{2}\Big([[\mathbf{M}_1,\mathbf{E}_1],\boldsymbol{\mu}_2]-[\mathbf{E}_1,[\mathbf{M}_1,\boldsymbol{\mu}_2]]
+[[\mathbf{M}_1,\mathbf{E}_0],\boldsymbol{\mu}_3]-[\mathbf{E}_0,[\mathbf{M}_1,\boldsymbol{\mu}_3]]\Big)\\
&= \frac{1}{2}\Big([[\mathbf{M}_1,\mathbf{E}_1],\boldsymbol{\mu}_2]-[\mathbf{E}_1,\mathbf{M}_2]
-2[\mathbf{E}_0,\mathbf{M}_3]
+[\mathbf{E}_0,[\mathbf{M}_2,\boldsymbol{\mu}_2]]
\Big)
\end{align}
\end{subequations}
Using now the relation \eqref{eq:[E,M]k} for $k=2$ on the first term, we can write
\begin{subequations}
	\begin{align}
		[\mathbf{M}_1,\mathbf{E}_2]
		&= \frac{1}{2}\Big(-[[\mathbf{M}_2,\mathbf{E}_0],\boldsymbol{\mu}_2]-[\mathbf{E}_1,\mathbf{M}_2]
		-2[\mathbf{E}_0,\mathbf{M}_3]
		+[\mathbf{E}_0,[\mathbf{M}_2,\boldsymbol{\mu}_2]]
		\Big)\\
				&= \frac{1}{2}\Big(-[\mathbf{M}_2,[\mathbf{E}_0,\boldsymbol{\mu}_2]]-[\mathbf{E}_1,\mathbf{M}_2]
		-2[\mathbf{E}_0,\mathbf{M}_3]
		\Big)\,.
	\end{align}
\end{subequations}
Finally, using the definition \eqref{eq:defE} of $\mathbf{E}_1$, we can write
	\begin{align}
	[\mathbf{M}_1,\mathbf{E}_2]
	&= -[\mathbf{E}_1,\mathbf{M}_2]
	-[\mathbf{E}_0,\mathbf{M}_3]
	\,,
\end{align}
which gives the relation \eqref{eq:[E,M]k} for $k=3$.

\subsection{Proof of \eqref{eq:diffE1}}
\label{app:proofdiff}

Here we will show that the recursive definition \eqref{eq:conjE} implies the differential equation \eqref{eq:diffE1} for the generating function $\mathbf{E}(t)$ given in \eqref{eq:Egen}. Starting with the right-hand side of \eqref{eq:diffE1}, we have
\begin{subequations}
	\begin{align}
		\text{RHS}[\eqref{eq:diffE1}]&=\sum_{k=0}^\infty\sum_{l=0}^\infty t^{k+l}[\mathbf{E}_k,\bs{\mu}_{l+2}]\\
		&=\sum_{m=1}^\infty t^{m-1}\sum_{k=0}^{m-1} [\mathbf{E}_k,\bs{\mu}_{m+1-k}]\,,
	\end{align}
\end{subequations}
where we have merely reordered the double sum in the second step. Recognizing the right-hand side of the recursive relation \eqref{eq:conjE}, we can finally write
\begin{subequations}
	\begin{align}
		\text{RHS}[\eqref{eq:diffE1}]
		&=\sum_{m=1}^\infty mt^{m-1}\mathbf{E}_m\\
 		&=\frac{\p}{\p t}\mathbf{E}(t)\\[4mm]
 		&=\text{LHS}[\eqref{eq:diffE1}]\,,
	\end{align}
\end{subequations}
which therefore gives the desired result.

\endgroup
\end{document}